 \documentclass[final,5p,times,twocolumn]{elsarticle}
\usepackage{hyperref}
\usepackage{amsmath}
\usepackage{amssymb}
\usepackage{amsthm}
\usepackage{xcolor}

\usepackage{amssymb}
\usepackage{amsmath}

\newcommand{\cb}{\color{black}}
\newcommand{\crb}[1]{\textcolor{black}{#1}}

\newcommand{\bi}{\begin{itemize}}

\renewcommand{\i}{\item}
\newcommand{\ei}{\end{itemize}}
\newcommand{\bd}{\begin{description}}
\newcommand{\ed}{\end{description}}
\newcommand{\be}{\begin{enumerate}}
\newcommand{\ee}{\end{enumerate}}
\newtheorem{thm}{Theorem}

\newtheorem{definition}{Definition}
\newtheorem{cor}{Corollary}
\newtheorem{rem}{Remark}
\newtheorem{lem}{Lemma}

\DeclareMathOperator{\rnd}{rnd}
\usepackage{amsthm}

\newcommand{\bqn}{\begin{eqnarray}}
\newcommand{\eqn}{\end{eqnarray}}
\newcommand{\eqnn}{\nonumber\end{eqnarray}}
\newcommand{\eqnl}[1]{\label{#1}\end{eqnarray}}
\newcommand{\bga}{\begin{gather}}
\newcommand{\ega}{\end{gather}}

\newcommand{\ba}[1]{\begin{array}{#1}}
\newcommand{\ea}{\end{array}}

\newcommand{\R}{\mathbb{R}}

\newcommand{\Z}{\mathbb{Z}}

\newcommand{\lam}{\lambda}

\newcommand{\om}{\omega}
\newcommand{\Om}{\Omega}

\newcommand{\sg}{\sigma}

\newcommand{\dt }{\dot}

\usepackage{xcolor}
\newcommand{\beq}{\begin{equation}}
\newcommand{\eeq}{\end{equation}}

\newcommand{\bag}{\begin{aligned}}
\newcommand{\eag}{\end{aligned}}

\usepackage[english]{babel}

 \makeatother

\usepackage{tikz}
\usepackage{pgfplots}
\usepackage{subfigure}
\usepackage{tikz-3dplot}
\usetikzlibrary{calc,arrows,shapes,backgrounds,calc,positioning,patterns,decorations.pathmorphing,decorations.markings,mindmap,trees,fadings,angles,quotes}
\usepackage{blox}
\usepackage{color}
\usepackage{booktabs}
\usepackage{balance}
\usepackage{xcolor}

\usepackage{dsfont}

 \newcommand{\E}{\mathrm{e}} 
  
\newtheorem{ass}{Assumption}

\usepackage{amsmath} 
\usepackage{graphicx} 
\usepackage{caption} 
\journal{}

\begin{document}

\begin{frontmatter}



\title{On hyperexponential stabilization of a chain of integrators in
continuous and discrete time  subject to  unmatched perturbations} 

\author[mymainaddress]{Moussa Labbadi\corref{mycorrespondingauthor}}
\cortext[mycorrespondingauthor]{Corresponding author}
\ead{moussa.labbadi@lis.lab.fr}

\author[mysecondaryaddress]{Denis Efimov}

\address[mymainaddress]{Aix-Marseille University, LIS UMR CNRS 7020, 13013 Marseille, France}
\address[mysecondaryaddress]{Inria, Univ. Lille, CNRS, UMR 9189 - CRIStAL, F-59000 Lille, France}


\begin{abstract}
A recursive time-varying state feedback is presented for a chain of integrators with unmatched perturbations in
continuous and discrete time. In continuous time, it is shown that hyperexponential convergence is achieved for the first state variable \(x_1\), while the second state \(x_2\) remains bounded. For the other states, we establish ISS {\cb property} by saturating the growing {\cb control} gain. In discrete time, we use implicit Euler discretization to {\cb preserve} hyperexponential convergence. The main results  are demonstrated through several examples of the proposed control laws, illustrating the conditions established  for both continuous and discrete-time systems.
\end{abstract}




\end{frontmatter}



\section{Introduction}

The problem of state feedback design, which ensures the asymptotic stability of the origin for continuous or discrete-time linear and nonlinear dynamical systems, is a cornerstone of control engineering. This foundational topic has been extensively studied, forming the basis of many modern control theories and techniques \cite{Chernousko2008, Isidori1995, Khalil2002, Utkin1992}. 

State feedback control methods are designed to achieve desired performance characteristics for closed-loop dynamics. Key considerations include:
\begin{itemize}
    \item[(i)] \textbf{Convergence rate}: {\cb Usually, control} laws can ensure exponential, finite-time, or fixed-time convergence of system trajectories to the origin. Finite-time control guarantees convergence within a bounded time dependent of initial conditions \cite{bhat2000finite}, while fixed-time control {\cb provides} convergence within a pre-specified duration regardless of the initial state \cite{polyakov2012nonlinear}.  
    \item[(ii)] \textbf{Robustness to disturbances}: Robustness is analyzed using the input-to-state stability (ISS) framework, which quantifies the influence of external disturbances on system trajectories \cite{dashkovskiy2011input, sontag2007input}. This property is essential for maintaining performance under uncertainties and perturbations.  
    \item[(iii)] \textbf{Robustness to mismatched  perturbations}: Persistent disturbances, not aligned with the control input direction, present  {\cb significant} challenges for robust control design {\cb \cite{Chernousko2008, Isidori1995}}. Effective strategies must attenuate such perturbations while minimizing control effort.  
    \item[(iv)] \textbf{Discrete-time implementation}: Implementing nonlinear feedback laws in discrete-time is challenging, especially for highly nonlinear stabilizers \cite{efimov2017robust, polyakov2019global}. In sliding mode control, the chattering phenomenon adds further complexity \cite{shtessel2014sliding}. Advanced methods are required to preserve stability and performance in the sampled-data domain.  
\end{itemize}

Accelerated convergence to equilibrium is crucial in many applications \cite{efimov2021phase} and is typically achieved via sliding mode or finite/fixed-time control laws. However, handling mismatched perturbations while ensuring accelerated convergence presents greater challenges compared to traditional exponential, finite-time, or fixed-time convergence with matched perturbations. 

Mismatched disturbances, which act on channels other than the control input, are common in various applications \cite{kayacan2019feedback, li2014continuous, moreno2021arbitrary, sun2014non, yang2013continuous, zhang2018nonsmooth}. Existing methods, such as backstepping, compensate for state-dependent uncertainties \cite[Sec. 7.1.2]{krstic1995nonlinear} or guarantee boundedness via ISS propagation through integrators \cite[Cor. 2.3]{jiang1994smallgain}. Other approaches leverage time-scaling or singular perturbation techniques \cite{praly1997generalized, ye2022robust, chitour2020stabilization, Khalil2002}. While these methods achieve asymptotic convergence \cite{praly1997generalized, Khalil2002} or prescribed-time convergence \cite{ye2022robust, chitour2020stabilization}, they do not fully address mismatched disturbances \cite{chitour2020stabilization} or assume state-dependent perturbations \cite{ye2022robust}.  

Recently, prescribed-time feedback control has emerged as a promising approach for linear and nonlinear systems, where time-varying gains grow unbounded within a fixed duration \cite{holloway2019time, song2017time, song2018time}. This method ensures uniform convergence in the presence of matched perturbations, akin to sliding mode control, but struggles to address mismatched disturbances {\cb and measurement noises}. Alternative concepts, such as hyperexponential convergence \cite{nekhoroshikh2022hyperexponential, wang2024exact, chu2022hyper}, provide  that equilibrium is reached asymptotically at a rate faster than any exponential dynamics.

For instance, hyperexponential convergence for the double integrator system was studied in \cite{efimov2022exact}, and extended to $n$th-order integrator chains with matched perturbations in \cite{wang2024exact, wang2024hyperexponential}, {\cb (in \cite{chu2022hyper} the disturbance-free case was studied)} using the following control law:
\[
    u(t) = K\Psi(t)x(t), \quad \Psi(t) = \text{diag}\{[\psi^n(t), \dots, \psi(t)]^\top\},\]
where \( K \in \mathbb{R}^{1 \times n} \) is the controller gain, and \(\psi(t) = 1 + t\) is a strictly increasing function of time. By tuning \( K \), the system achieves uniform stability with hyperexponential convergence, surpassing the speed of conventional exponential methods.   However, in the presence of unmatched uncertainties, this technique fails to maintain stability. Additionally, due to the complexity of stability analysis and feedback construction, it remains unclear how to quantitatively evaluate the limitations of this approach or extend its applicability.

Motivated by the above discussion, this paper extends the study of a hyperexponentially  stabilizing feedback for a double integrator subjected to mismatched perturbations, as proposed in \cite{labbadi2024hyperexponential}, where  uniform convergence with respect to both matched and mismatched disturbances was demonstrated; however, the implementation in discrete time was not addressed.   In this paper, we generalize these results to an arbitrary-order chain of integrators, which represents a more complex system requiring {\cb an extended} analysis. While the double integrator case was addressed in two-step, the general case necessitates a more sophisticated theoretical framework and specific conditions for time-varying gains.  In continuous time, we prove that hyperexponential convergence is achieved for the first state variable, \( x_1 \), while ensuring that the second state, \( x_2 \), remains bounded. For higher-order states, we establish input-to-state stability  by saturating the growth of the gain \( \psi \).  In discrete time, we employ an implicit Euler discretization scheme to {\cb maintain the} hyperexponential convergence for  state variables. This approach preserves the key properties of the continuous-time counterpart while also ensuring boundedness for the entire state vector.  Furthermore, we present several illustrative examples of the proposed control law, demonstrating its stability and effectiveness for both continuous- and discrete-time systems.  

 \section*{Notation}
 \begin{itemize}
     \item \( \mathbb{R}^+ = \{x \in \mathbb{R} : x \ge 0\} \), where \( \mathbb{R} \) represents the set of real numbers, \( \mathbb{Z} \) denotes the set of integers, and \( \mathbb{Z}^+ = \mathbb{Z} \cap \mathbb{R}^+ \) signifies the set of non-negative integers.
    
     \item The absolute value in \( \mathbb{R} \) is denoted by \( | \cdot | \), and \( \| \cdot \| \) represents the Euclidean norm on \( \mathbb{R}^n \).
    
     \item For a (Lebesgue) measurable function \( d : \mathbb{R}^+ \to \mathbb{R}^m \), the norm \( \|d\|_\infty = \text{ess sup}_{t \ge 0} \|d(t)\| \) is defined. The set of functions \( d \) satisfying \( \|d\|_\infty < +\infty \) is denoted as \( L^\infty_m \).
    
      \item For a sequence \( d_k \in \mathbb{R}^m, \; k \in \mathbb{Z}^+ \), define the norm 
\[
|d|_\infty = \sup_{k \in \mathbb{Z}^+} \| d_k \|,
\]
and the set of sequences with \( | \cdot |_\infty < +\infty \) is further denoted as \( {\cb \ell}^\infty_m \).
    
     \item A continuous function \( \alpha : \mathbb{R}^+ \to \mathbb{R}^+ \) belongs to the class \( \mathcal{K} \) if \( \alpha(0) = 0 \) and it is strictly increasing. A function \( \alpha : \mathbb{R}^+ \to \mathbb{R}^+ \) is in the class \( \mathcal{K}_\infty \) if \( \alpha \in \mathcal{K} \) and \( \alpha \) increases to infinity. A continuous function \( \beta : \mathbb{R}^+ \times \mathbb{R}^+ \to \mathbb{R}^+ \) is in the class \( \mathcal{KL} \) if \( \beta(\cdot, t) \in \mathcal{K} \) for each fixed \( t \in \mathbb{R}^+ \) and \( \beta(s, \cdot) \) decreases to zero for each fixed \( s >0 \). 
     \item The identity matrix of dimension \( n \times n \) is denoted by \( I_n \).
     \item $\exp(1)= \E$.
 \end{itemize}

\section{Preliminaries}
The standard stability concepts for non-autonomous nonlinear systems whose dynamics described by ordinary differential equations, as the asymptotic stability and the input-to-state stability, can be found in \cite{Khalil2002}. Below, less conventional notions are recalled together with the auxiliary lemma used in the paper.
\subsection{Uniform hyperexponential stability}
\begin{enumerate}
\item {\cb Continuous-time case: } Consider the following non-autonomous system:
\begin{equation}\label{eq:sdfn}
\frac{dx(t)}{dt} = f(t, x(t), d(t)), \quad t \geq t_0, \quad t_0 \in \mathbb{R}^+
\end{equation}
where \( x(t) \in \mathbb{R}^n \) is the state, \( d(t) \in \mathbb{R}^m \) represents the external disturbances, \( d \in L^\infty_m \), and \( f : \mathbb{R}^+ \times \mathbb{R}^n \times \mathbb{R}^m \to \mathbb{R}^n \) is a continuous function with respect to \( x \) and \( d \), and piecewise continuous with respect to \( t \). Additionally, \( f(t, 0, 0) = 0 \) for all \( t \in \mathbb{R}^+ \). The solution of system \eqref{eq:sdfn} with an initial condition \( x_0 \in \mathbb{R}^n \) at time \( t_0 \in \mathbb{R}^+ \) and a disturbance \( d \in L^\infty_m \) is denoted as \( \varphi(t, t_0, x_0, d) \). {\cb We assume that \( f \) ensures the existence and uniqueness of solutions \( \varphi(t, t_0, x_0, d) \) in forward time.}
The following definition introduces the uniform stability property for \eqref{eq:sdfn} with a hyperexponential rate of convergence.
\begin{definition} \cite{wang2024exact,wang2024hyperexponential} 
\crb{The system \eqref{eq:sdfn} is said to be} uniformly hyperexponentially stable if there exist constant \( \kappa_0 \in \mathbb{R}^+ \), functions \( \rho, \kappa \in \mathcal{K}_\infty \) and \( \beta \in \mathcal{KL} \) such that:
\begin{equation}\label{eq:s_hyp}
\bag
\|\varphi(t, t_0, x_0, d)\| \leq \E^{-(\kappa(t-t_0)+\kappa_0)(t-t_0)}\rho(\|x_0\|) + \beta(\|d\|_\infty, t - t_0),
 \quad \qquad \qquad  \forall t \geq t_0
\eag
\end{equation}
for all \( x_0 \in \mathbb{R}^n \), \( t_0 \in \mathbb{R}^+ \), and \crb{\( d \in L_m^\infty \)}.   
\crb{It is said to be uniformly hyperexponentially stable in the variable $x_1$ if
$$\|\varphi_1(t, t_0, x_0, d)\|\leq\E^{-(\kappa(t-t_0)+\kappa_0)(t-t_0)}\rho(\|x_0\|) + \beta(\|d\|_\infty, t - t_0),
 \quad \qquad \qquad  \forall t \geq t_0$$
for some $\kappa_0\in\R^+$, $\rho,\kappa\in\mathcal{K}_\infty$, $\beta\in\mathcal{KL}$ and for all $x_0\in\R^n$, $t_0\in\R^+$, $d\in L_m^\infty$.}
\end{definition}

An illustrative example is:
\begin{equation}\label{eq:example1}
\dot{x}(t) = -(1 + t)x(t) + d(t), \quad t \geq 0
\end{equation}
where \( x(t), d(t) \in \mathbb{R} \). Let $t_0 = 0$, the solutions to  \eqref{eq:example1} can be bounded by:
\begin{equation}
|x(t)| \leq \E^{-(\frac{t^2}{2} + t)}|x(0)| + \frac{2\|d\|_\infty}{1 + (\frac{t^2}{2} + t)}
\end{equation}
for all \( t \geq  0 \), any \( x(0) \in \mathbb{R} \), and \( d \in L^\infty_1 \). Thus,  we can define \( \rho(s) = s \), \( \kappa(s) = \frac{s}{2} \), \( \kappa_0 = 1 \), and \( \beta(s, t) = \frac{2s}{1 + \left(\frac{t}{2} + 1\right)t} \).

In this definition, the hyperexponential rate of convergence is required only for the initial conditions, whereas uniformity is understood as independence from  the exogenous input \( d \in \crb{L_1^\infty} \) and the initial time $t_0$. Although \crb{\( d \in L^\infty_m \)} is assumed, any other suitable class of inputs can be considered. 

\item{\cb Discrete-time case:  
 Consider the nonautonomous difference equation:
\begin{equation}\label{eq:system_DT}
x_{k+1} = f(k, x_k, d_k), \quad k \geq k_0, \quad k_0 \in \mathbb{Z}_+,
\end{equation}
where $x_k \in \mathbb{R}^n$ is the state vector, $d_k \in \mathbb{R}^m$ is the disturbance vector with $d \in \ell_\infty^m$, $f: \mathbb{Z}_+ \times \mathbb{R}^n \times \mathbb{R}^m \to \mathbb{R}^n$ satisfies $f(k, 0, 0) = 0$ for all $k \in \mathbb{Z}_+$. The solution for initial condition $x_0 \in \mathbb{R}^n$ at $k_0 \in \mathbb{Z}_+$ and an input $d\in\ell_\infty^m$ is denoted by $X_k(k_0, x_0, d)$, which we also assume uniquely defined for $k\in\Z_+$.
\begin{definition}\label{def:uhs} \cite{wang2024hyperexponential} 
\crb{The system \eqref{eq:system_DT} is said to be} uniformly hyperexponentially stable if $\exists \rho, \kappa \in \mathcal{K}_\infty$, $\kappa_0 \in (0, \kappa^{-1}(1))$, and $\beta \in \mathcal{KL}$ such that:
\begin{equation}\label{eq:stability}
\|X_k(k_0, x_0, d)\| \leq \kappa(\kappa_0 + k - k_0)^{k_0 - k}\rho(\|x_0\|) + \beta(|d|_\infty, k - k_0)
\end{equation}
holds for all $ k \geq k_0$, $x_0 \in \mathbb{R}^n$, $k_0 \in \mathbb{Z}_+$, and $d \in \ell_\infty^m$.
\end{definition}
For an illustration, consider a scalar system \cite{wang2024hyperexponential}:
\begin{equation}\label{eq:example2}
x_{k+1} = \frac{x_k + d_k}{1 + k}, \quad k \in \mathbb{Z}_+,
\end{equation}
with $x_k, d_k \in \mathbb{R}$. Its solution satisfies:
\begin{align}
|x_k| \leq \frac{|x_0| + |d_0|}{k!} + \frac{1}{k}\sum_{i=1}^{k-1}|d_i|\prod_{j=i+1}^{k-1}\frac{1}{j} 
\leq \frac{|x_0|}{k!} + \frac{\|d\|_\infty}{k}\sum_{i=1}^k \frac{1}{2^{i-2}} \label{eq:sol2}
\end{align}
for all $k \geq 1$, $\forall x_0 \in \mathbb{R}$, $d \in \ell_\infty^1$.
This system is uniformly hyperexponentially stable with:
 $\rho(s) = s$, $\kappa(s) = \frac{s}{2}$, $\kappa_0 = 1$ (using Stirling's approximation $k! \geq \sqrt{2\pi k}(\frac{k}{\E})^k$), $\beta(s,k) = \frac{8s}{1+k^2}$ when $k_0 = 0$.
 }
\end{enumerate}
\subsection{Auxiliary properties}
In this paper, the following lemma will be used.
\begin{lem}\label{lem:Update_3}\cite{labbadi2024hyperexponential}
For all \( \tau \geq 0 \), and $a,\alpha>0$ such that \( \alpha a > 1 \), the following holds:
\begin{gather*}
\int_{0}^{\tau}\E^{s-\tau}\frac{ds}{(a s + 1)^\alpha} \leq \frac{r_{a,\alpha}}{(a \tau + 1)^\alpha},
\end{gather*}
where
\begin{gather*}
r_{a,\alpha} = (a \alpha)^{\alpha} \int_{0}^{\frac{a\alpha - 1}{a}} \E^{s - \frac{a\alpha - 1}{a}} \frac{ds}{(a s + 1)^{\alpha}}  \\ + (a \alpha + 1) + \left(\frac{a \alpha + 1}{a \alpha}\right)^{\alpha} \left(1 - \E^{-\frac{1}{a}}\right).
\end{gather*}
\end{lem}
\section{Problem statement}
Consider a linear system in a Brunovsk\'{y}-regular canonical form with matched and mismatched perturbations.
\begin{equation}\label{eq:sysm-0} 
\begin{aligned}
\dot{x}(t) = &\, e_n x(t) + d(t) + b_n  u(t), \quad t \geq 0,
\end{aligned}
\end{equation}
\[b_n = \begin{bmatrix}
0_{(n-1) \times 1} \\ 1
\end{bmatrix}, \quad e_n = \begin{bmatrix}
0_{(n-1) \times 1} & I_{n-1} \\
0 & 0_{1 \times (n-1)}
\end{bmatrix}, \]
where $x(t) \in \mathbb{R}^n$ is the state vector, $u(t) \in \mathbb{R}$ is the control, $d(t) = [d_1(t), \dots, d_n(t)]^\top \in \mathbb{R}^n$ denotes matched and mismatched uncertainties.   The admissible perturbations considered in this paper satisfy the following assumption.  
\begin{ass}\label{assumption:perturbations}
\crb{
The disturbance $d$ is essentially bounded.}
\end{ass}
The objective is to design a  state feedback control \( u(t) = u(x(t), t) \) for this nonlinear system, which ensures stabilization of \( x(t) \) at the origin with a hyperexponential convergence rate, {\cb robustly} in \( d   \). Any controllable nonlinear single-input, single-output system can be represented in this canonical form, with controllability serving as a key requirement for achieving accelerated stabilization.

In the following sections, we will first address the problem in continuous time. Then, we will extend the approach to its discrete-time counterpart.

\section{Control in continuous time}
Let the auxiliary variables  be chosen recursively in a linear time-varying form:
\beq\label{eq:s}
\begin{aligned}
 \sg_1(t) &= x_1(t),\\
 \sg_2(t) &= \om_1(t) + \lambda_{1} \sg_1(t) \psi(t),\\
 \sigma_3(t) &= \om_2(t) + \lambda_{2} \sg_2(t) \psi^{2}(t),\\
 & \vdots \\
 \sigma_{n}(t) &= \om_{n-1}(t) + \lambda_{n-1} \sigma_{n-1}(t) \psi^{n-1}(t),
\end{aligned}
\eeq
where  \( \lam_i, i=1,\dots,n-1 \) are tuning gains such that \( \lam_{i+1} > \lam_i \), $\psi(t)=1+t$ and each recursive term $\sigma_i(t)$ is constructed to reject unmatched perturbations by progressively increasing the  gains  
 \cite{alessandri2013time,holloway2019time, efimov2024discretization}. The term $\om_i(t)$ represents the dynamics of each $\sg_i(t)$ such that $\om_i(t) = \dot \sg_i(t)$ under the condition $d \equiv 0$. For instance, we have $\om_1(t) = x_2(t)$ because $\dot \sg_1(t) = \dot x_1(t) = x_2(t) + d_1(t)$, where $d_1 \equiv 0$ is considered for design of auxiliary variables. Similarly, $\om_2(t) = x_3(t) + \lambda_1\left(x_1(t) + x_2(t)\psi(t)\right)$ follows from $\dot \sg_2(t) = x_3(t) + d_2(t) + \lambda_1\left(x_1(t) + (x_2(t) + d_1(t))\psi(t)\right)$, given that $d_1 = d_2 \equiv 0$. We can use the same steps to determine the expressions for all $\om_i(t)$. 
 
We introduce the  control law for \eqref{eq:sysm-0} as follows:
\beq \label{eq:u}
 u(t)  = -\Om(t,x(t)) - \lam_n \sg_{n}(t) \psi^m(t),
\eeq
{\cb where  \( \lam_n > \lam_{n-1}\), \( m \geq n \), and \(\Omega(t, x(t))\) represents the nominal system dynamics that must be canceled to achieve the desired behavior of the highest-order auxiliary variable \(\sigma_n(t)\) in the absence of perturbations. This corresponds to the last step of the recursive design procedure, where the control input \(u(t)\) directly influences the highest-order state. Specifically, \(\Omega(t, x(t))\) includes all known terms from the derivative of \(\sigma_n(t)\) under the assumption \(d \equiv 0\), such that the control law cancels these terms and imposes the desired dynamics through the term \(-\lambda_n \sigma_n(t) \psi^m(t)\).
For example, for the double integrator system, \(\Omega(t, x(t))\) is given by
\[
\Omega(t, x(t)) = -\lambda_1 \big( \psi(t) x_2(t) + x_1(t) \big).
\]
For the third-order integrator, \(\Omega(t, x(t))\) takes the form
\begin{gather*}
\Omega(t, x(t)) = -\big[ 2 \lambda_1 + (2 + \lambda_1 \psi^2(t)) \lambda_2 \psi(t) \big] x_2(t)\\ 
- 3 \lambda_2 \lambda_1 \psi^2(t) x_1(t)
- \psi(t) (\lambda_1 + \lambda_2 \psi(t)) x_3(t).
\end{gather*}
}
 For the matched case, similar control approaches were used for prescribed-time stabilization in \cite{chitour2020stabilization} and hyperexponential stabilization in \cite{wang2024hyperexponential} with  function $\psi(t)$ that escapes to infinity in finite time \cite{chitour2020stabilization}, {\cb or a similar}  one  in \cite{wang2024hyperexponential}.

\begin{rem}
Any continuous, strictly increasing function 
\(\psi : \mathbb{R}^+ \to \mathbb{R}^+\) with \(\psi(0) > 0\) and an unbounded integral 
can be utilized in \eqref{eq:u}. 
For instance, \(\psi(t) = a \E^{\alpha t}\) or \(\psi(t) = b t^t\), 
where \(a > 0\), \(b > 0\), and \(\alpha > 0\). \hfill $\triangle$
\end{rem}
\begin{rem} Prescribed-time convergence can be achieved using functions of the form
\(
\psi(t) = \frac{\psi_0}{T - t},
\)
as proposed in~\cite{chitour2020stabilization, holloway2019time, efimov2024discretization, ye2022robust}, 
where \(T > 0\) is the predefined settling time and \(\psi_0 > 0\). 
\crb{However, since they are not well-defined for all \(t \ge 0\) and exhibit a singularity as \(t \to T^-\), 
they are excluded from the admissible class of \(\psi\) in the main result statements.} 
As discussed in~\cite{aldana2023inherent}, systems based on this approach are highly sensitive to even arbitrarily small \crb{measurement} perturbations, 
which leads to significant robustness issues.
\hfill $\triangle$
\end{rem}

To establish the stability of the closed-loop system under {\cb \eqref{eq:u}}, a new time variable should be defined to facilitate the derivation of the general solution to the closed-loop system  \eqref{eq:sysm-0}, beginning with the  variable $\sg_n(t)$. 

\begin{thm}\label{theo:n}
Consider the system \eqref{eq:sysm-0} under Assumption \ref{assumption:perturbations}, 
controlled by the time-varying algorithm \eqref{eq:u}. 
If the tuning gains  $\lambda_{i+1}$ \crb{are} sufficiently larger than $\lambda_i$, 
then the state component \(x_1\) converges to zero with a uniform hyperexponential rate, 
while the component \(x_2\) \crb{remains bounded}. 
Moreover, if the disturbances $d_i$, $i=1,\dots,n-1$, are $n-i$ times continuously differentiable 
\crb{with bounded derivatives}, then the closed-loop system is ISS 
\crb{with respect to the disturbance and its derivatives}.
\end{thm}
\crb{
\begin{cor}\label{cor:1}
For any $\epsilon > 0$ and any disturbance $d$ satisfying $\|d\|_\infty \leq \Delta$ for a given $\Delta > 0$, 
there exists a constant $\Psi_{\epsilon,\Delta} > 0$ such that replacing in (\ref{eq:s}), (\ref{eq:u}) 
the gain $\psi(t)$ by $\min\{\psi(t), \Psi_{\epsilon,\Delta}\}$ ensures that the closed-loop system is ISS. 
Moreover, for every initial condition $x(0) \in \mathbb{R}^n$, there exists a time 
$T_{\epsilon,\Delta}(x(0)) > 0$ such that 
\[
|x_1(t)| \leq \epsilon, \quad \forall t \geq T_{\epsilon,\Delta}(x(0)).
\]
\end{cor}
}
The {\cb outcomes} claimed in the theorem are as follows:
\bi
    \i For \( d_i, i = 1, \dots, n-1 \) being sufficiently differentiable \cite{deng2024robust}, we can perform a direct change of coordinates:
\begin{gather*}
\tilde{x}_2 = x_2 + d_1, \quad \tilde{x}_3 = x_3 + d_2 + \dt  d_1, \quad \dots, \\
 \quad \tilde{x}_n = x_n + d_{n-1} + \dt d_{n-2} + \dots + \frac{d^{n-1}}{dx^{n-1}} d_1.
\end{gather*}
This change of coordinates allows us to reduce the original problem to the regulation of a chain of integrators model with matched disturbances. In this new formulation, the perturbation \( d_n \) as well as all time derivatives of \( d_i, i = 1, \dots, n-1 \) are   appearing in the last equation only. The theorem guarantees uniform hyperexponential stability of the origin in the closed-loop system described by equations \eqref{eq:sysm-0}-\eqref{eq:u} under these transformations.
However, the theorem's results require a sufficient smoothness of $d$ with existence of  upper bound for all derivatives.
 
    \i For \( x_i \) with \( i = 3, \dots, n \), if the growth of \( \psi(t) \) becomes saturated once \( |x_1| \leq \epsilon \) for some specified design parameter \( \epsilon \), then the asymptotic behavior of these variables has finite gains in \crb{disturbances}{\cb, while the control signal $u(t)$ stays bounded}. Similar results, based on the saturation of time-varying and increasing gains{\cb, have} been obtained in \cite{alessandri2013time, alessandri2015increasing}.
\ei

 \begin{proof}
  Consider the  dynamics of \eqref{eq:s} that can be written  as follows:
\begin{align}\label{eq:dt_s}
\begin{aligned}
 \dot \sg_1(t) & = x_2(t) + d_1(t),\\
 \dot \sg_2(t) & = \dot \om_1(t) + \lam_1 \left( \dot \sg_1(t) \psi(t) + \sg_1(t) \right),\\
 \dot \sg_3(t) & = \dot \om_2(t) + \lam_2 \left( \dot \sg_2(t) \psi^{2}(t) + 2\psi(t) \sg_2(t) \right),\\
 & \vdots \\ 
\dot \sg_{n-2}(t) & = \dot \om_{n-3}(t) + \lam_{n-3} \left( \dot \sg_{n-3}(t) \psi^{n-3}(t)  + (n-3) \psi^{n-4}(t) \sg_{n-3}(t) \right),\\ 
 \dot \sg_{n-1}(t) & = \dot \om_{n-2}(t) + \lam_{n-2} \left( \dot \sg_{n-2}(t) \psi^{n-2}(t)  + (n-2) \psi^{n-3}(t) \sg_{n-2}(t) \right),\\
 \dot \sg_{n}(t) & = \dot \om_{n-1}(t) + \lam_{n-1} \left( \dot \sg_{n-1}(t) \psi^{n-1}(t)  + (n-1) \psi^{n-2}(t) \sg_{n-1}(t) \right).
\end{aligned}
\end{align}

 As discussed after \eqref{eq:s}, the term $\om_i(t)$ does not depend on perturbations; however, its derivative is influenced by the perturbations, rather than by the derivatives of the perturbations themselves. Moreover, each $\dt \om_i(t)$ depends on $\dt x_{i+1}(t)$.

 Now, skipping for brevity the dependence of all variables on time, we have:
\begin{align}\label{eq:dt_s2}
\begin{aligned}
 \dt \sg_{n}  &= \dt \om_{n-1} + (n-1) \lam_{n-1} \psi^{n-2} \sg_{n-1} + \lam_{n-1} \psi^{n-1} \dt \sg_{n-1}\\ 
 &= \dt \om_{n-1} + (n-1) \lam_{n-1} \psi^{n-2} \sg_{n-1} + \lam_{n-1} \psi^{n-1} 
 \dt \om_{n-2} \\ & \qquad   
  + (n-2) \lam_{n-1}\lam_{n-2} \psi^{2n-4} \sg_{n-2} 
+ \lam_{n-1} \lam_{n-2} \psi^{2n-3} \dt \sg_{n-2}\\
 &= \dt \om_{n-1} + (n-1) \lam_{n-1} \psi^{n-2} \sg_{n-1} + \lam_{n-1} \psi^{n-1} 
 \dt \om_{n-2}    \\ & \qquad  
  + (n-2) \lam_{n-1}\lam_{n-2} \psi^{2n-4} \sg_{n-2} + \lam_{n-1} \lam_{n-2} \psi^{2n-3} \dot \om_{n-3}\\
& \quad  + (n-3)\lam_{n-1} \lam_{n-2}\lam_{n-3} \psi^{3n-7}\sg_{n-3} 
\\
& \quad + \lam_{n-1} \lam_{n-2}\lam_{n-3} \psi^{3n-6} 
\dot \sg_{n-3}. 
\end{aligned}
\end{align}

Consequently, to obtain $\dt \sg_n$, we must recursively substitute all instances of the time derivatives of $\sg_{i-1}$ recalling that the term $\omega_{i-1} = \dt \sg_{i-1}$ when $d_i \equiv 0$, and we should replace all time derivatives of $\sg_{i-1}$ to derive the corresponding expression. The final time derivative  is influenced by the input $u$ and all perturbations, including both matched and mismatched components. This derivative consists of two parts: a known component, which is related to the auxiliary variables, and an unknown component, which is a function proportional to the perturbations $d$. The known part simplifies to the function $\Omega$, while the remaining terms contribute to the expression for $\dt \sg_n$. 

For instance, in the case of a second-order system, there is only one term $\om_1$, whose derivative is expressed as $\dt \om_1 = \dt x_2 = u  + d_2$. By substituting the control input $u = -( \Om + \lam_2 \psi^2 \sg_2)$, we obtain the time derivative of $\om_1$ as $\dt \om_1 = \dt x_2 = - \Omega - \lam_2 \psi^2 \sg_2 + d_2${\cb, where $\Omega= -\lam_1(\psi x_2+x_1)$}. 

Next, considering the equation for $\dt \sg_2$, we have $\dt \sg_2 = \dt \om_1 + \lam_1 x_1 + \lam_1 \psi x_2 + \lam_1 \psi d_1 = - \Omega - \lam_2 \psi^2 \sg_2 + d_2 + \lam_1 x_1 + \lam_1 \psi x_2 + \lam_1 \psi d_1$. Here, the term $ \Om = \lam_1(x_1 + \psi x_2)$, which simplifies the expression for $\dt \sg_2$ to:
\[
\dt \sg_2 = - \lam_2 \psi^2 \sg_2 + d_2 + \lam_1 \psi d_1.
\]
From this example, it can be seen that $\dt \sg_n$ depends directly on the matched perturbation $d_n$, while the mismatched perturbations are scaled by the time-varying function $\psi$ and the tuning parameters $\lam_i$. Then, we can represent the time derivative of $\sg_n$ in the closed-loop system as follows:
 \[
\begin{aligned}
\dot{\sg}_n &= -\lam_n \psi^m \sg_n + d_n + P_1^n(\psi) d_{n-1} + P_2^n(\psi) d_{n-2}  + P_3^n(\psi) d_{n-3}\\
& \qquad + \dots + P_{n-2}^n(\psi)d_2  + P_{n-1}^n(\psi)d_1,
\end{aligned}
\]
where \( P_1^n, P_2^n, \dots, P_{n-1}^n \) are polynomials in \( \psi \) with degrees \( 1, 2, \dots, n-1 \), respectively, and their coefficients depend on \( \lambda_i \). The same structure applies to the other variables:
\begin{align*}\label{eq:dt_vf}
\begin{aligned} \dt \sg_1 & = -\lam_1 \psi \sg_1+\sg_2 + d_1,\\
 \dot \sg_2 & = -\lam_2 \psi^2\sg_2 + d_2 +\sg_3 + P_1^2(\psi)d_1,\\
 \dot \sg_3  &=  -\lam_3 \psi^3\sg_3 + d_3+\sg_4  + P_1^3(\psi)d_2 + P_2^3(\psi)d_1,\\ 
\dot \sg_{n-2} & = -\lam_{n-2} \psi^{n-2}\sg_{n-2} + d_{n-2} + P_1^{n-2}(\psi) d_{n-3}  
 + P_2^{n-2}(\psi) d_{n-4}\\ & \qquad +\sg_{n-1}+ P_3^{n-2}(\psi) d_{n-5} + \dots + P_{n-4}^{n-2}(\psi)d_2   + P_{n-3}^{n-2}(\psi)d_1,\\
 \dot \sg_{n-1}  & = -\lam_{n-1} \psi^{n-1}\sg_{n-1} + d_{n-1}+\sg_{n} + P_1^{n-1}(\psi) d_{n-2} \\ & \qquad  
 + P_2^{n-1}(\psi) d_{n-3} + P_3^{n-1}(\psi) d_{n-4} + \dots + P_{n-3}^{n-1}(\psi)d_2 \\ & \qquad  
  + P^{n-1}_{n-2}(\psi)d_1.
\end{aligned}
\end{align*}
This formulation provides a clear structure, showing the dependency of each \(\dot{\sg}_i\) on both matched and mismatched perturbations \(d_i\), and detailing how the polynomials \(P_i^j(\psi)\) progressively incorporate these terms with increasing degrees. The stabilization process is sequential, beginning with \(\sg_n\) and progressing down to \(\sg_1 = x_1\). 

To achieve stability for \(x_1 =\sigma_1 \), we utilize \(\sg_2\), then we employ \(\sg_3\) to stabilize \(\sigma_2\), continuing this cascade effect down to \(\sg_n\). {\cb On each recursive step,} \(\sg_i\) plays a role in controlling the subsequent variable \(\sg_{i-1}\), ensuring a robust overall stabilization of the entire system.

Let us introduce a new time variable as:
 \begin{gather*}
d\tau = \lam_n \psi^m(t) \, dt \implies\\
 \tau = \varphi(t) = \frac{\lam_n}{m+1} \left( b_1 t + b_2 t^2 + \dots + b_{m+1}t^{m+1} \right), \\
t = \varphi^{-1}(\tau) = \sqrt[m+1]{(m+1)\lam_n^{-1}\tau + 1} - 1,\quad\\
  \text{and} \quad \psi(\tau) = \sqrt[m+1]{(m+1)\lam_n^{-1}\tau + 1},
\end{gather*}
where \( b_j \) are positive coefficients determined based on \( m \) according to specific known identities.

The variable $\sg_i$ and the disturbances in the new time frame can then be written as:
     \[
     \sg_i(\tau) = \sg_i\left[\varphi^{-1}(\tau)\right], \ d_i(\tau) = d_i\left[\varphi^{-1}(\tau)\right], \ i=1,\dots, n.
     \]
     We have:
  \begin{gather*}
     \frac{d}{d\tau} \sg_n(\tau) = \frac{d\sg_n(t)}{dt} \bigg|_{t = \varphi^{-1}(\tau)} \frac{d\varphi^{-1}(\tau)}{d\tau}\\
     = - \sg_n(\tau) 
     +  \frac{\lam_n^{-1} d_n(\tau)}{ \psi^m(\tau)} +  \frac{\lam_n^{-1} P_1^n(\psi(\tau)) d_{n-1}(\tau)}{ \psi^m(\tau)} \\  + 
     \frac{\lam_n^{-1} P_2^n(\psi(\tau)) d_{n-2}(\tau)}{ \psi^m(\tau)} +
\frac{\lam_n^{-1} P_3^n(\psi(\tau)) d_{n-3}(\tau)}{ \psi^m(\tau)}\\+ \dots
+\frac{\lam_n^{-1} P_{n-2}^n(\psi(\tau)) d_2(\tau)}{ \psi^m(\tau)}             
 + \frac{\lam_n^{-1} P_{n-1}^n (\psi(\tau)) d_1(\tau)}{ \psi^m(\tau)}.
 \end{gather*}

    By transitioning to the time domain, we obtain the following solution:
      \begin{gather*}
     \sg_n(\tau) = \E^{-\tau} \sg_n(0) + \int_0^\tau \E^{s-\tau}\Bigg(  \frac{\lam_n^{-1} P_1^n(\psi(s)) d_{n-1}(s)}{\psi^m(s)} +\frac{\lam_n^{-1} d_n(s)}{\psi^m(s)} \\  + 
     \frac{\lam_n^{-1} P_2^n(\psi(s)) d_{n-2}(s)}{\psi^m(s)}
\frac{\lam_n^{-1} P_3^n(\psi(s)) d_{n-3}(s)}{\psi^m(s)}+ \dots\\
+\frac{\lam_n^{-1} P_{n-2}^n(\psi(s)) d_2(s)}{\psi^m(s)}             
 + \frac{\lam_n^{-1} P_{n-1}^n (\psi(s)) d_1(s)}{\psi^m(s)} \Bigg) ds.
     \end{gather*}
    
    The upper bound of \( \sg_n \) can be derived as:
\begin{gather*}
|\sg_n(\tau)| \leq \E^{-\tau} |\sg_n(0)| + \int_0^\tau \E^{s-\tau}\Bigg(  
 \frac{\lam_n^{-1} \|d_n\|_\infty}{\psi^m(s)}\\ + \frac{\lam_n^{-1} P_1^n(\psi(s)) \|d_{n-1}\|_\infty}{\psi^m(s)} 
+ \frac{\lam_n^{-1} P_2^n(\psi(s)) \|d_{n-2}\|_\infty}{\psi^m(s)} \\+ \frac{\lam_n^{-1} P_3^n(\psi(s)) \|d_{n-3}\|_\infty}{\psi^m(s)} 
+ \dots\\ + \frac{\lam_n^{-1} P_{n-2}^n(\psi(s)) \|d_2\|_\infty}{\psi^m(s)} + \frac{\lam_n^{-1} P_{n-1}^n(\psi(s)) \|d_1\|_\infty}{\psi^m(s)} \Bigg) ds.
\end{gather*}
By substituting the $P_i^j$ expression, yields:
\begin{gather*}
|\sg_n(\tau)| \leq \E^{-\tau} |\sg_n(0)| + \int_0^\tau \E^{s-\tau}\Bigg(  
 \frac{\lam_n^{-1} \|d_n\|_\infty}{\psi^m(s)} \\ + \frac{\lam_n^{-1}\bigg( a_{10} +a_{11}\psi(s)\bigg) \|d_{n-1}\|_\infty}{\psi^m(s)} 
\\+ \frac{\lam_n^{-1} \bigg(a_{20} +a_{21}\psi(s)  +a_{22}\psi^2(s)) \bigg)\|d_{n-2}\|_\infty}{\psi^m(s)} 
+ \dots\\ + \frac{\lam_n^{-1} \bigg(a_{(n-1)0}+  \dots +a_{(n-1)n-1}\psi^{n-1}(s)\bigg) \|d_1\|_\infty}{\psi^m(s)} \Bigg) ds,
\end{gather*}
where $a_i$ are non-negative coefficients. 
    After some simple calculations, we get:
    
\begin{gather*}
|\sg_n(\tau)| \leq \E^{-\tau} |\sg_n(0)| + \int_0^\tau \E^{s-\tau}\Bigg(  
 \frac{\lam_n^{-1} \|d_n\|_\infty}{\psi^m(s)}+ \frac{\lam_n^{-1} a_{10} \|d_{n-1}\|_\infty}{\psi^m(s)} \\ +\frac{\lam_n^{-1}a_{11} \|d_{n-1}\|_\infty}{\psi^{m-1}(s)} 
+ \frac{\lam_n^{-1} a_{20}\|d_{n-2}\|_\infty}{\psi^m(s)}  +\frac{\lam_n^{-1} a_{21}\|d_{n-2}\|_\infty}{\psi^{m-1}(s)}  \\ + \frac{\lam_n^{-1} a_{22}\|d_{n-2}\|_\infty}{\psi^{m-2}(s)} 
+ \dots + \frac{\lam_n^{-1} a_{(n-1)0}\|d_1\|_\infty}{\psi^m(s)}\\+  \dots +\frac{\lam_n^{-1}a_{(n-1)n-1} \|d_1\|_\infty}{\psi^{m-n+1}(s)} \Bigg) ds.
\end{gather*}    
    
 {\cb By applying Lemma \ref{lem:Update_3} with the coefficient \( r_{a,\alpha} \)  defined as }\( a = \frac{m}{\lam_n} \) and \( \alpha = \left\{\frac{m}{m+1}, \frac{m-1}{m+1}, \frac{m-2}{m+1},\dots, \frac{m-n+1}{m+1}\right\} \) we have: 
 \begin{gather*}
|\sg_n(\tau)| \leq \E^{-\tau} |\sg_n(0)| + 
\frac{\lam_n^{-1} r_{\frac{\lam_n}{m},\frac{m}{m+1}}\|d_n\|_\infty}{\psi^m(\tau)} 
+ \frac{\lam_n^{-1}r_{\frac{\lam_n}{m},\frac{m}{m+1}} a_{10} \|d_{n-1}\|_\infty}{\psi^m(\tau)}\\
 + \frac{\lam_n^{-1} r_{\frac{\lam_n}{m},\frac{m-1}{m+1}} a_{11} \|d_{n-1}\|_\infty}{\psi^{m-1}(\tau)} 
+ \frac{\lam_n^{-1}r_{\frac{\lam_n}{m},\frac{m}{m+1}} a_{20}\|d_{n-2}\|_\infty}{\psi^m(\tau)}\\ + \frac{\lam_n^{-1} r_{\frac{\lam_n}{m},\frac{m-1}{m+1}} a_{21}\|d_{n-2}\|_\infty}{\psi^{m-1}(\tau)} + \frac{\lam_n^{-1}r_{\frac{\lam_n}{m-2},\frac{m}{m+1}} a_{22}\|d_{n-2}\|_\infty}{\psi^{m-2}(\tau)} \\
+ \dots 
+ \frac{\lam_n^{-1}r_{\frac{\lam_n}{m},\frac{m}{m+1}} a_{(n-1)0}\|d_1\|_\infty}{\psi^m(\tau)} \\+ \dots  + \frac{\lam_n^{-1} r_{\frac{\lam_n}{m},\frac{m-n+1}{m+1}}a_{(n-1)(n-1)} \|d_1\|_\infty}{\psi^{m-n+1}(\tau)}.
\end{gather*}

Returning to the original time frame:
\begin{gather*}
|\sg_n(t)| \leq \E^{-\frac{\lam_n}{m+1} \left( b_1 t + b_2 t^2 + \dots + t^{m+1} \right)} |\sg_n(0)| + 
\frac{\lam_n^{-1} r_{\frac{\lam_n}{m},\frac{m}{m+1}}\|d_n\|_\infty}{\psi^m(t)} \\ + \frac{\lam_n^{-1}r_{\frac{\lam_n}{m},\frac{m}{m+1}} a_{10} \|d_{n-1}\|_\infty}{\psi^m(t)}
 + \frac{\lam_n^{-1} r_{\frac{\lam_n}{m},\frac{m-1}{m+1}} a_{11} \|d_{n-1}\|_\infty}{\psi^{m-1}(t)} \\
+ \frac{\lam_n^{-1}r_{\frac{\lam_n}{m},\frac{m}{m+1}} a_{20}\|d_{n-2}\|_\infty}{\psi^m(t)} + \frac{\lam_n^{-1} r_{\frac{\lam_n}{m},\frac{m-1}{m+1}} a_{21}\|d_{n-2}\|_\infty}{\psi^{m-1}(t)}\\  + \frac{\lam_n^{-1}r_{\frac{\lam_n}{m-2},\frac{m}{m+1}} a_{22}\|d_{n-2}\|_\infty}{\psi^{m-2}(t)} 
+ \dots \\
+ \frac{\lam_n^{-1}r_{\frac{\lam_n}{m},\frac{m}{m+1}} a_{(n-1)0}\|d_1\|_\infty}{\psi^m(t)} + \dots  + \frac{\lam_n^{-1} r_{\frac{\lam_n}{m},\frac{m-n+1}{m+1}}a_{(n-1)(n-1)} \|d_1\|_\infty}{\psi^{m-n+1}(t)}\\{\cb 
\leq \E^{-\frac{\lambda_n}{m+1} \left( b_1 t + b_2 t^2 + \dots + t^{m+1} \right)} |\sigma_n(0)| 
+ \lambda_n^{-1} \sum_{k=0}^{n-1} \sum_{j=0}^{k} \frac{r_{\frac{\lambda_n}{m-j}, \frac{m-j}{m+1}} \, a_{kj} \, \|d_{n-k}\|_\infty}{\psi^{m-j}(t)}.}
\end{gather*}
with \( m > n - 1 \), this condition imposes an upper bound on \( \sg_n \), ensuring that the variable remains bounded and undergoes a hyperexponential decay relative to its initial conditions. The imposed decay rate is substantially enhanced by the disturbance compensation terms, resulting in a rapid and robust convergence of  \( \sg_n \) to zero. 


{\cb For the other variables $\sigma_i$ ($n-1,\dots,2$), we can use the same time scale transformations as follows:
\begin{equation}
d\tau_i = \lambda_i \psi^{f(m,i)}(t)\,dt, \quad \text{where} \quad f(m,i) = m - (n -1 - i)
\end{equation}
with the following explicit form:
\begin{equation}
\begin{cases}
\sigma_{n-1} &: d\tau_{n-1} = \lambda_{n-1} \psi^{m-1}(t)\,dt \\
\sigma_{n-2} &: d\tau_{n-2} = \lambda_{n-2} \psi^{m-2}(t)\,dt \\
&\vdots \\
\sigma_i &: d\tau_i = \lambda_i \psi^{m-(n-i)}(t)\,dt \\
&\vdots \\
\sigma_2 &: d\tau_2 = \lambda_2 \psi^{m-(n-2)}(t)\,dt 
\end{cases}
\end{equation}
and the time-scale $\tau_i$:
\begin{align*}
\tau_i &= \varphi_i(t) = \frac{\lambda_i}{f(m,i)+1}\left(b_1 t + b_2 t^2 + \cdots + b_{f(m,i)+1} t^{f(m,i)+1}\right), \\
t &= \varphi_i^{-1}(\tau_i) = \sqrt[f(m,i)+1]{(f(m,i)+1)\lambda_i^{-1}\tau_i + 1} - 1, \\
\psi(\tau_i) &= \sqrt[f(m,i)+1]{(f(m,i)+1)\lambda_i^{-1}\tau_i + 1}.
\end{align*}

When $m = n$, the transformations simplify since $f(n,i) = n - (n - i) +1 = i+1$.
}

After showing the boundedness of all auxiliary variables \crb{$\sg_i, i=1,\dots,n$}, \crb{the analysis proceeds to the state variables.} We begin with the equation
\[
\dot{x}_1 = -\lambda_1 x_1 \psi + d_1 + \sg_2,
\]
and apply the following time scale transformation:
\[
d\tau = \lambda_1 \psi(t) \, dt.
\]
\crb{Under this change, the dynamics of \(x_1\) become
\[
\frac{dx_1}{d\tau} = -x_1 + \frac{d_1 + \sigma_2}{\lambda_1 \psi}.
\]
Integration gives
\[
x_1(\tau) = x_1(0) \, \E^{-\tau} + \int_0^\tau \E^{-(\tau-s)} \frac{d_1(s) + \sigma_2(s)}{\lambda_1 \psi(t(s))} \, ds.
\]
Let \(\crb{C_1(\tau) = \frac{M_1}{\lambda_1} \int_0^\tau \E^{-(\tau-s)} \frac{1}{\psi(t(s))} \, ds}\), where 
\(M_1 = \sup_{s \geq 0} |d_1(s) + \sigma_2(s)| < \infty\). Then \(\lim_{\tau \to \infty} C_1(\tau) = 0\).  
Hence,
\[
|x_1(t)| \leq |x_1(0)| \, \E^{-\frac{\lambda_1 t}{2} (2+t)} + C_1(t),
\]
showing that \(x_1(t)\) converges to zero, with hyperexponential convergence rate in initial conditions.}



For the state \(x_2\), using the relation
\[
x_2 = \sigma_2 - \lambda_1 \psi(t) x_1,
\]
\crb{we have
\[
|x_2| \le |\sigma_2| + \lambda_1 \psi |x_1| \le |\sigma_2| + \lambda_1 \psi \left( |x_1(0)| \, \E^{-\frac{\lambda_1 t}{2} (2+t)} + C_1 \right).
\]}
Let \crb{\(C_2 = \lambda_1 C_1 \psi\)}, which converges to a finite constant as \(t \to \infty\).  
Since \(\sigma_2(t)\) converges hyperexponentially and the term \crb{\(C_2\)} remains bounded, it follows that \(x_2(t)\) is bounded.

We can apply the same process for all variables. As \crb{we can conclude from these computations}, the asymptotic behavior of the variables \( x_i \) with \( i = 3, \dots, n \) is proportional to \( d_{i-2}, \dots, d_1 \) with the respective gains \( \psi, \dots, \psi^{i-2} \). {\cb Therefore,}

\begin{itemize}
    \i If the growth of \( \psi(t) \) is saturated, as given in Corollary \ref{cor:1}, this implies the boundedness of these variables.

    \i If the mismatched \( d_i \) are sufficiently smooth, then due to the boundedness and continuity of all variables, and the convergence of \( x_1(t) \) to zero, it follows that
    \[
    \dot{x}_1(t) \to 0 \quad \text{as} \quad t \to +\infty,
    \]
that implies convergence of $x_2(t)$ to $-d_1(t)$.
    Hence, repeating the same arguments as \( t \to +\infty \):
    \begin{gather*}
    x_2(t) = -d_1(t),\; 
    x_3(t) = -d_2(t) - \frac{d}{dt} d_1(t), \;
    \cdots,\; \\
    x_n(t) = -d_{n-1}(t) - \frac{d}{dt} d_{n-2}(t) - \dots - \frac{d^{n-1}}{dt^{n-1}} d_1(t).
\end{gather*}
\end{itemize}
This completes the proof.
 \end{proof}
 
 \section{Special cases}
 {\cb In this subsection we examplify the results of Theorem~\ref{theo:n} and Corollary~\ref{cor:1} to the particular values of $n$.}

 \subsection{Second-order system}
We consider the following double integrator:
\begin{align}
\begin{aligned}
    \dot{x}_1(t) &= x_2(t) + d_1(t), \\
    \dot{x}_2(t) &= u(t) + d_2(t), \label{eq:perturbed_integrator}
\end{aligned}
\end{align}
Auxiliary variables are given as \cite{labbadi2024hyperexponential}:
\[
    \sg_1(t) = x_1(t), \quad \sg_2(t) = x_2(t) + \lambda_1 \psi(t) x_1(t),
\]
and control law as \cite{labbadi2024hyperexponential}:
\begin{equation} \label{eq:u_sos}
    u(t) = -\lambda_1\left(\psi(t) x_2(t) + x_1(t)\right) - \lambda_2 \psi^2(t) \sg_2(t),
\end{equation}
where \( \lambda_1 \) and \( \lambda_2 \) are control gains.

The following result is presented for the double integrator:
\crb{\begin{cor}\label{theo:bounded}\cite{labbadi2024hyperexponential}
Consider the system \eqref{eq:perturbed_integrator} under Assumption~\ref{assumption:perturbations}, 
controlled by \eqref{eq:u_sos}. If $\lambda_2 > \tfrac{3}{2}\lambda_1$, then the closed-loop system is ISS and uniformly hyperexponentially stable with respect to \( x_1 \). 
If, in addition, $d_1$ is continuously differentiable with bounded derivative, then for every initial 
condition $x(0)\in\mathbb{R}^2$ it holds that
\(
\lim_{t\to\infty} x_2(t) = -d_1(t).
\)
\end{cor}
}
This result demonstrates  the specified conditions presented in Theorem \ref{theo:n} for $n=2$.

 \subsection{Third-Order System}
The results presented in the previous section for second-order systems will now be extended to the following third-order system:
\begin{align}
\begin{aligned}
	\dot{x}_1(t) &= x_2(t) + d_1(t), \\
	\dot{x}_2(t) &= x_2(t) + d_2(t), \\
	\dot{x}_3(t) &= u(t) + d_3(t). \label{eq:third_integrator}
\end{aligned}
\end{align}

By leveraging the formulation from the second-order system, we can now define  additional auxiliary variable as follows:
\begin{gather}
\sg_1(t) = x_1(t), \quad \sg_2(t) = x_2(t)+ \lambda_1\psi(t) x_1(t), \nonumber \\
\dt \sg_2 = -\lam_2\psi^2\sg_2 +d_2 +\lambda_1 \psi d_1, \nonumber \\
\sg_3(t) = x_3(t)+ \lam_1\left(\psi(t) x_2(t)+x_1\right) +\lam_2\psi^2(t)\sg_2(t) \label{eq:s_third}\\
 =  x_3(t)+ \left(\lam_2\psi(t) +\lam_1\right) \psi(t)x_2(t)+\lam_1\left(\lam_2\psi^3(t)+1\right)x_1(t)\nonumber.
\label{eq:s_thirly}
\end{gather}

{\cb Control (\ref{eq:u}) takes the following form in this case for $m=4$:}
\beq \label{eq:u_thirtly}
\bag
	u(t) =& -\lam_3\psi^4(t)\sg_3(t) -\left[2\lam_1+(2+\lam_1\psi^2(t))\lam_2\psi(t)\right]x_2(t)\\
	&-  3\lam_2\lam_1\psi^2(t)x_1(t) -\psi(t)(\lam_1+\lam_2\psi(t))x_3(t).
	\eag
\eeq
{\cb Theorem~\ref{theo:n} admits the following reformulation for the third order integrator:}
\crb{
\begin{thm}\label{theo:third}
Consider the system \eqref{eq:third_integrator}--\eqref{eq:u_thirtly} under Assumption~\ref{assumption:perturbations}. 
If $\lambda_2 > \tfrac{3}{2}\lambda_1$ and $\lambda_3 > \tfrac{5}{3}\lambda_2$, then the system is uniformly hyperexponentially stable in the variable $x_1$. 
 In addition, $x_2$ remains bounded for all $t\geq 0$.
\end{thm}
}
\crb{
\begin{cor}
For any $\epsilon>0$ and $\Delta>0$ with $\|d\|_\infty \leq \Delta$, there exists a constant $\Psi_{\epsilon,\Delta} > 0$ such that, if the gain $\psi(t)$ in \eqref{eq:s_third} and \eqref{eq:u_thirtly} is replaced by
\(
\min\{\psi(t), \Psi_{\epsilon,\Delta}\},
\)
then the closed-loop system is ISS, and the state component $x_1(t)$ satisfies
\[
|x_1(t)| \le \epsilon \quad \text{for all } t \ge T_{\epsilon,\Delta}(x(0)),
\] 
for every initial condition $x(0) \in \mathbb{R}^3$.
\end{cor}
}

\begin{proof}
	We begin by noting the following:
\begin{gather*}
	\dot{\sg}_3 = 3\lam_3\lam_1 \psi^2 x_1 + \left[2\lam_1 + (2 + \lam_1 \psi^2)\lam_2\psi\right] x_2\\
 + \psi(\lam_1 + \lam_2\psi)x_3 + u+ d_3
	 + \lam_1 \left(\lam_2\psi^3 + 1\right) d_1  + \psi(\lam_1 + \lam_2\psi)d_2\\
	= -\lam_3\psi^4(t) \sg_3 + \lam_1 \left(\lam_2\psi^3 + 1\right) d_1 + \psi(\lam_1 + \lam_2\psi(t))d_2 + d_3.
\end{gather*}

 	Next, we introduce a new time variable  as follows:
\begin{gather*}
	d\tau = \lam_3 \psi^4(t) dt, \implies\\
	\tau = \varphi(t) = \lam_3 t \left(\frac{t^4}{5} + t^3 + 2t^2 + 2t + 1\right),\\
	t = \varphi^{-1}(\tau) = \sqrt[5]{5\lam_3^{-1} \tau + 1} - 1,
	\psi(\tau) = \sqrt[5]{5\lam_3^{-1} \tau + 1}.
\end{gather*}

	The variables in the new time frame can then be written as:
	\begin{gather*}
	\sg_3(\tau) = \sg_3\left[\varphi^{-1}(\tau)\right], \quad d_i(\tau) = d_i\left[\varphi^{-1}(\tau)\right], \quad i=1,2,3,
	\end{gather*}
	yielding the following expression:
 \begin{gather*}
    \frac{d}{d\tau} \sg_3(\tau) = \frac{d\sg_3(t)}{dt} \bigg|_{t = \varphi^{-1}(\tau)} \frac{d\varphi^{-1}(\tau)}{d\tau} \\
     = - \sg_3(\tau) + \lam_1 \lam_3^{-1} \left[\frac{\lam_2}{\psi(\tau)} + \frac{1}{\psi^4(\tau)}\right] d_1(\tau)\\    
      + \lam_3^{-1}\left[\frac{\lam_1}{\psi^3(\tau)} + \frac{\lam_2}{\psi^2(\tau)}\right] d_2(\tau)  + \frac{\lam_3^{-1} d_3(\tau)}{\psi^4(\tau)}.
 \end{gather*}

	By transitioning to the time domain, we obtain the following exact solution:
\begin{gather*}
	\sg_3(\tau) = \E^{-\tau}\sg_3(0) + \int_0^\tau \E^{s-\tau} \Bigg( \lam_1 \lam_3^{-1}\Bigg[\frac{\lam_2}{\psi(s)}  + \frac{1}{\psi^4(s)}\Bigg]  d_1(s)
	\\ + \lam_3^{-1}\left[\frac{\lam_1 }{\psi^3(s)} + \frac{\lam_2}{\psi^2(s)}\right] d_2(s)
	 + \frac{\lam_3^{-1} d_3(s)}{\psi^4(s)} \Bigg) ds.
\end{gather*}
    
	The upper bound of \( \sg_3 \) can be derived as:
\begin{gather*}
	|\sg_3(\tau)| \leq \E^{-\tau} |\sg_3(0)| + \int_0^\tau \E^{s-\tau}
 \Bigg( \lam_1 \lam_3^{-1} \left[\frac{\lam_2}{\psi(s)} + \frac{1}{\psi^4(s)}\right] \|d_1\|_\infty
	\\ + \lam_3^{-1}\left[\frac{\lam_1 }{\psi^3(s)} + \frac{\lam_2}{\psi^2(s)}\right] \|d_2\|_\infty
	+ \frac{\lam_3^{-1} \|d_3\|_\infty}{\psi^4(s)} \Bigg) ds.
\end{gather*}
    
	{\cb By applying Lemma \ref{lem:Update_3},  where the coefficient \( r_{a,\alpha} \) is defined with \( a = \frac{5}{\lam_3} \) and \( \alpha = \left\{\frac{1}{5}, \frac{4}{5}, \frac{3}{5}, \frac{2}{5}\right\} \), and r}eturning to the original time:
\begin{gather*}
	|\sg_3(t)| \leq \\ 
    \E^{-\lam_3 t \left(\frac{t^4}{5} + t^3 + 2t^2 + 2t + 1\right)} |\sg_3(0)|
	 + \lam_1 \lam_3^{-1} \left[\frac{\lam_2 r_{\frac{5}{\lam_3},\frac{1}{5}}}{\psi(t)} + \frac{r_{\frac{5}{\lam_3},\frac{4}{5}}}{\psi^4(t)}\right] \|d_1\|_\infty\\ + \frac{\lam_3^{-1}r_{\frac{5}{\lam_3},\frac{4}{5}} }{\psi^4(t)}\|d_3\|_\infty
	 + \lam_3^{-1}\left[\frac{\lam_1 r_{\frac{5}{\lam_3},\frac{3}{5}}}{\psi^3(t)} + \frac{\lam_2 r_{\frac{5}{\lam_3},\frac{2}{5}}}{\psi^2(t)}\right] \|d_2\|_\infty.
\end{gather*}

	This upper bound on  $\sg_3$ indicates that the {\cb variable} is bounded and experiences a hyperexponential decay relative to initial conditions. The decay is significantly accelerated by the applied compensation for  disturbances, ensuring rapid and robust convergence of the system's state to the surface $\sg_3 =0 $.

Next, we analyze the dynamics of $\sg_2(t)$ as
\begin{gather*}
\dot{\sg}_2(t) = x_3 + d_2 + \lam_1\left(\psi x_2 + x_1\right) + \lam_1 \psi d_1 \\
            	= -\lam_3 \psi^2 \sg_2 + d_2 + \lam_1 \psi d_1 + \sg_3.
\end{gather*}
To facilitate the analysis, let us introduce a new time variable, defined as:
\begin{gather*}
d\tau = \lam_2 \psi^2(t) dt \implies
\tau = \varphi(t) = \lam_2 t \left(\frac{t^2}{3} + t + 1\right),\\
t = \varphi^{-1}(\tau) = \sqrt[3]{3\lam_2^{-1}\tau + 1} - 1,
\psi(\tau) = \sqrt[3]{3\lam_2^{-1}\tau + 1}.
\end{gather*}
Thus, the  variables in the new time can be expressed as:
\begin{gather*}
\sg_2(\tau) = \sg_2\left[ \varphi^{-1}(\tau) \right], \quad \sg_3(\tau) = \sg_3\left[ \varphi^{-1}(\tau) \right], \\
d_i(\tau) = d_i\left[ \varphi^{-1}(\tau) \right].
\end{gather*}
Consequently, we obtain:
\begin{gather*}
\frac{d}{d\tau}\sg_2(\tau) = \frac{d\sg_2(t)}{dt} \bigg|_{t = \varphi^{-1}(\tau)} \frac{d\varphi^{-1}(\tau)}{d\tau}\\
 = - \sg_2(\tau) +\frac{\lam_2^{-1}d_2(\tau)}{\left(\sqrt[3]{3 \lam_2^{-1} \tau +1}\right)^2} + \frac{\lam_1 \lam_2^{-1} d_1(\tau)}{\sqrt[3]{3 \lam_2^{-1} \tau +1}}
 +\frac{\lam_2^{-1}\sg_3(\tau)}{\left(\sqrt[3]{3 \lam_2^{-1} \tau +1}\right)^2}.
\end{gather*}
Upon transitioning to the new time domain, a solution  for $\sg_2(\tau)$ can be derived as follows:
\begin{gather*}
\sg_2(\tau) = \E^{-\tau} \sg_2(0) +\\ 
 \lam_2^{-1}\int_0^\tau\E^{s-\tau}\Bigg[\frac{d_2(s)}{\psi^2(s)} + \frac{\lam_1  d_1(s)}{\psi(s)}
 +\frac{\sg_3(s)}{\psi^2(s)} \Bigg]ds.
\end{gather*}
We can now establish the upper bound of $\sg_2(\tau)$ as:
\begin{gather*}
|\sg_2(\tau)| \leq \E^{-\tau} |\sg_2(0)|
+\lam_2^{-1}\int_0^\tau\E^{s-\tau}\Bigg[\frac{\|d_2\|_\infty}{\psi^2(s)} + \frac{\lam_1  \|d_1\|_\infty}{\psi(s)}
+\frac{|\sg_3(s)|}{\psi^2(s)} \Bigg]ds.
\end{gather*}
We further refine this bound as:
\begin{gather*}
|\sg_2(\tau)| \leq \E^{-\tau} |\sg_2(0)|+
 \lam_2^{-1}\int_0^\tau\E^{s-\tau}\Bigg[\frac{\|d_2\|_\infty}{\psi^2(s)}  + \frac{\lam_1  \|d_1\|_\infty}{\psi(s)}
\\+\frac{\E^{-\frac{\lam_3}{5}\left[\left(\sqrt[3]{\frac{3s}{\lam_2} +1}\right)^2(\frac{3s}{\lam_2} +1)-1\right]} |\sg_3(0)|}{\psi^2(s)} \\
	 + \lam_1 \lam_3^{-1}\left[\frac{\lam_2 r_{\frac{5}{\lam_3},\frac{1}{5}}}{\psi^3(s)} + \frac{r_{\frac{5}{\lam_3},\frac{4}{5}}}{\psi^6(s)}\right] \|d_1\|_\infty + \frac{\lam_2^{-1}\lam_3^{-1} r_{\frac{5}{\lam_3},\frac{4}{5}}\|d_3\|_\infty}{\psi^6(s)}\\
	 + \lam_3^{-1}\left[\frac{\lam_1 r_{\frac{5}{\lam_3},\frac{3}{5}}}{\psi^5(s)} + \frac{\lam_2r_{\frac{5}{\lam_3},\frac{2}{5}}}{\psi^4(s)}\right] \|d_2\|_\infty\Bigg]ds,
\end{gather*}
where the additional terms account for contributions from the disturbances $d_1$, $d_2$, and $d_3$. {\cb Using the following estimate:}
\begin{gather*}
\int_0^\tau\E^{s-\tau}\frac{\E^{-\frac{\lam_3}{5}\left[\left(\sqrt[3]{\frac{3s}{\lam_2} +1}\right)^2(\frac{3s}{\lam_2} +1)-1\right]}}{\lam_2 \left(\sqrt[3]{\frac{3s}{\lam_2} +1}\right)^2}ds
\leq \int_0^\tau\E^{s-\tau}\frac{\E^{-\frac{3\lam_3 }{ 5\lam_2}s}}{\lam_2}ds \\  = 5  \frac{\E^{-\tau} -\E^{-\frac{3\lam_3}{5\lam_2}\tau} }{3\lam_3 - 5\lam_2} \leq   \frac{5\E^{-\tau}}{3\lam_3 - 5\lam_2},
\end{gather*}
and applying Lemma \ref{lem:Update_3}, the upper bound on   $\sg_2$ satisfies:
\begin{gather*}
|\sg_2(\tau)| \leq \E^{-\tau} |\sg_2(0)| +\frac{5\E^{-\tau}}{3\lam_3 - 5\lam_2}|\sg_3(0)| \\
	 + \frac{\lam_1}{\lam_3\lam_2} \Bigg[\frac{\lam_2 r_{\frac{5}{\lam_3},\frac{1}{5}} r_{\frac{3}{\lam_2},1}}{\psi^3(\tau)} + \frac{r_{\frac{5}{\lam_3},\frac{4}{5}} r_{\frac{3}{\lam_2},2}}{\psi^6(\tau)}
	+ \frac{\lam_3 r_{\frac{3}{\lam_2},\frac{1}{3}}}{\psi(\tau)}\Bigg] \|d_1\|_\infty \\
	\\ + \frac{1}{\lam_2\lam_3}\Bigg[\frac{\lam_1 r_{\frac{5}{\lam_3},\frac{3}{5}} r_{\frac{3}{\lam_2},\frac{5}{3}}}{\psi^5(\tau)}  + \frac{\lam_2r_{\frac{5}{\lam_3},\frac{2}{5}} r_{\frac{3}{\lam_2},\frac{4}{3}}}{\psi^4(\tau)}
	+\frac{\lam_3r_{\frac{3}{\lam_2},\frac{2}{3}}}{\psi^2(\tau)}\Bigg] \|d_2\|_\infty \\
 	+ \frac{\lam_2^{-1}\lam_3^{-1} r_{\frac{5}{\lam_3},\frac{4}{5}} r_{\frac{3}{\lam_2},2}}{\psi^6(\tau)}\|d_3\|_\infty.
\end{gather*}
Returning to the original time:
\begin{gather*}
|\sg_2(t)| \leq \E^{-\lam_2 t \left(\frac{t^2}{3} + t + 1\right)} |\sg_2(0)|
 +\frac{5\E^{-\lam_2 t \left(\frac{t^2}{3} + t + 1\right)} }{3\lam_3 - 5\lam_2}|\sg_3(0)| \\
	 + \frac{\lam_1}{\lam_3\lam_2} \Bigg[\frac{\lam_2 r_{\frac{5}{\lam_3},\frac{1}{5}} r_{\frac{3}{\lam_2},1}}{\psi^3(t)} + \frac{r_{\frac{5}{\lam_3},\frac{4}{5}} r_{\frac{3}{\lam_2},2}}{\psi^6(t)}
	+ \frac{\lam_3 r_{\frac{3}{\lam_2},\frac{1}{3}}}{\psi(t)}\Bigg] \|d_1\|_\infty \\
	\\ + \frac{1}{\lam_2\lam_3}\Bigg[\frac{\lam_1 r_{\frac{5}{\lam_3},\frac{3}{5}} r_{\frac{3}{\lam_2},\frac{5}{3}}}{\psi^5(t)}  + \frac{\lam_2r_{\frac{5}{\lam_3},\frac{2}{5}} r_{\frac{3}{\lam_2},\frac{4}{3}}}{\psi^4(t)}
	+\frac{\lam_3r_{\frac{3}{\lam_2},\frac{2}{3}}}{\psi^2(t)}\Bigg] \|d_2\|_\infty \\
 	+ \frac{\lam_2^{-1}\lam_3^{-1} r_{\frac{5}{\lam_3},\frac{4}{5}} r_{\frac{3}{\lam_2},2}}{\psi^6(t)}\|d_3\|_\infty,
\end{gather*}
where the decay of the  $\sg_2$ is hyperexponential. This indicates that this variable is bounded and exhibits rapid convergence, accelerated by disturbance compensation, ensuring robust system performance.

 On $\sg_2$, the dynamics of $x_1$ remain identical to those presented for {\cb the} second-order system. Consequently, we introduce a new time variable following the same formulation as provided in the second part of the proof in the preceding section. The upper bound of $x_1$ in new time   is given by:
\begin{gather*}
|x_1(\tau)| \leq \int_0^\tau\E^{s-\tau}\Bigg[\frac{\|d_1\|_\infty }{\lam_1\psi(s)}
 + \frac{\E^{-\frac{\lam_2}{3}\left[\sqrt{\frac{2s}{\lam_1} +1}(\frac{2s}{\lam_1} +1)-1\right]}}{\lam_1\psi(s)}|\sg_2(0)| \\ +\E^{-\tau} |x_1(0)| +\frac{5\E^{-\frac{\lam_2}{3}\left[\sqrt{\frac{2s}{\lam_1} +1}(\frac{2s}{\lam_1} +1)-1\right]}}{\lam_1(3\lam_3 - 5\lam_2)\psi(s)}|\sg_3(0)| \\ 
+ \frac{\lam_1}{\lam_3\lam_2} \Bigg[\frac{\lam_2 r_{\frac{5}{\lam_3},\frac{1}{5}} r_{\frac{3}{\lam_2},1}}{\psi^4(s)} + \frac{r_{\frac{5}{\lam_3},\frac{4}{5}} r_{\frac{3}{\lam_2},2}}{\psi^7(s)}
	+ \frac{\lam_3 r_{\frac{3}{\lam_2},\frac{1}{3}}}{\psi(s)}\Bigg] \|d_1\|_\infty 
	\\ + \frac{1}{\lam_2\lam_3}\Bigg[\frac{\lam_1 r_{\frac{5}{\lam_3},\frac{3}{5}} r_{\frac{3}{\lam_2},\frac{5}{3}}}{\psi^6(s)}  + \frac{\lam_2r_{\frac{5}{\lam_3},\frac{2}{5}} r_{\frac{3}{\lam_2},\frac{4}{3}}}{\psi^5(s)}
	+\frac{\lam_3r_{\frac{3}{\lam_2},\frac{2}{3}}}{\psi^3(s)}\Bigg] \|d_2\|_\infty \\ 
 	+ \frac{\lam_2^{-1}\lam_3^{-1} r_{\frac{5}{\lam_3},\frac{4}{5}} r_{\frac{3}{\lam_2},2}}{\psi^7(s)}\|d_3\|_\infty
 \Bigg]ds.
\end{gather*}
Use the following inequality:
\begin{gather*}
\int_0^\tau  \frac{\E^{s-\tau}\E^{-\frac{\lam_2}{3}\left[\sqrt{\frac{2s}{\lam_1} +1}(\frac{2s}{\lam_1} +1)-1\right]}}{\lam_1\sqrt{\frac{2s}{\lam_1} +1}}\, ds 
 \leq
 \int_0^\tau   \frac{\E^{s-\tau}\E^{-\frac{2\lam_2}{3\lam_1}s}}{\lam_1}\, ds \\
 =  3\frac{\E^{-\tau} - \E^{-\frac{2\lam_2}{3\lam_1}\tau}}{2\lam_2 -3\lam_1}\leq 3\frac{\E^{-\tau}}{2\lam_2 -3\lam_1},
\end{gather*}
with $2\lam_2>3\lam_1$.
According to Lemma \ref{lem:Update_3}, by applying the above  inequalities, the upper bound estimate of the state $x_1$ is obtained as follows:
\begin{gather*}
|x_1(\tau)| \leq \E^{-\tau} |x_1(0)|
 + \frac{3\E^{-\tau}}{2\lam_2 -3\lam_1}|\sg_2(0)| \\ + \frac{15\E^{-\tau}}{(3\lam_3 - 5\lam_2)(2\lam_2 -3\lam_1)}|\sg_3(0)| \\
+ \frac{1}{\lam_3\lam_2} \Bigg[\frac{\lam_2 r_{\frac{5}{\lam_3},\frac{1}{5}} r_{\frac{3}{\lam_2},1} r_{\frac{2}{\lam_1},2}}{\psi^4(\tau)} + \frac{r_{\frac{5}{\lam_3},\frac{4}{5}} r_{\frac{3}{\lam_2},2}r_{\frac{2}{\lam_1},\frac{7}{2}}}{\psi^7(\tau)}\\ 
	+ \frac{\lam_3 r_{\frac{3}{\lam_2},\frac{1}{3}}r_{\frac{2}{\lam_1},\frac{1}{2}}}{\psi(\tau)} +\frac{\lam_2\lam_3r_{\frac{2}{\lam_1},\frac{1}{2}}}{\lam_1\psi(\tau)}\Bigg] \|d_1\|_\infty \\
	 + \frac{\lam_1^{-1}}{\lam_2\lam_3}\Bigg[\frac{\lam_1 r_{\frac{5}{\lam_3},\frac{3}{5}} r_{\frac{3}{\lam_2},\frac{5}{3}}r_{\frac{2}{\lam_1},3}}{\psi^6(\tau)}  + \frac{\lam_2r_{\frac{5}{\lam_3},\frac{2}{5}} r_{\frac{3}{\lam_2},\frac{4}{3}}r_{\frac{2}{\lam_1},\frac{5}{2}}}{\psi^5(\tau)}
	\\+\frac{\lam_3r_{\frac{3}{\lam_2},\frac{2}{3}}r_{\frac{2}{\lam_1},\frac{3}{2}}}{\psi^3(\tau)}\Bigg] \|d_2\|_\infty 
 	+ \frac{\lam_2^{-1}\lam_3^{-1} r_{\frac{5}{\lam_3},\frac{4}{5}} r_{\frac{3}{\lam_2},2}r_{\frac{2}{\lam_1},\frac{7}{2}}}{\psi^7(\tau)}\|d_3\|_\infty.
\end{gather*}
And in the original time:
\crb{\begin{gather*}
|x_1(t)| \le A_1(t)\,|x_1(0)| + A_2(t)\,|\sigma_2(0)| + A_3(t)\,|\sigma_3(0)|\\ 
+ B_1(t)\,\|d_1\|_\infty + B_2(t)\,\|d_2\|_\infty + B_3(t)\,\|d_3\|_\infty,
\end{gather*}
where
\begin{align*}
A_1(t) &= \E^{-\lambda_1 \frac{t}{2}(t+2)}\quad
A_2(t) = \frac{3 \E^{-\lambda_1 \frac{t}{2}(t+2)}}{2 \lambda_2 - 3 \lambda_1}, \\
A_3(t) &= \frac{15 \E^{-\lambda_1 \frac{t}{2}(t+2)}}{(3 \lambda_3 - 5 \lambda_2)(2 \lambda_2 - 3 \lambda_1)}, \\
B_1(t) &= \frac{1}{\lambda_3 \lambda_2} \Bigg[
\frac{\lambda_2 r_{\frac{5}{\lambda_3},\frac{1}{5}} r_{\frac{3}{\lambda_2},1} r_{\frac{2}{\lambda_1},2}}{\psi^4(t)}
+ \frac{r_{\frac{5}{\lambda_3},\frac{4}{5}} r_{\frac{3}{\lambda_2},2} r_{\frac{2}{\lambda_1},\frac{7}{2}}}{\psi^7(t)}\\ & \qquad 
+ \frac{\lambda_3 r_{\frac{3}{\lambda_2},\frac{1}{3}} r_{\frac{2}{\lambda_1},\frac{1}{2}}}{\psi(t)}
+ \frac{\lambda_2 \lambda_3 r_{\frac{2}{\lambda_1},\frac{1}{2}}}{\lambda_1 \psi(t)}
\Bigg], \\
B_2(t) &= \frac{1}{\lambda_2 \lambda_3 \lambda_1} \Bigg[
\frac{\lambda_1 r_{\frac{5}{\lambda_3},\frac{3}{5}} r_{\frac{3}{\lambda_2},\frac{5}{3}} r_{\frac{2}{\lambda_1},3}}{\psi^6(t)}
+ \frac{\lambda_2 r_{\frac{5}{\lambda_3},\frac{2}{5}} r_{\frac{3}{\lambda_2},\frac{4}{3}} r_{\frac{2}{\lambda_1},\frac{5}{2}}}{\psi^5(t)}
\\ & \qquad  + \frac{\lambda_3 r_{\frac{3}{\lambda_2},\frac{2}{3}} r_{\frac{2}{\lambda_1},\frac{3}{2}}}{\psi^3(t)}
\Bigg], \quad
B_3(t) = \frac{r_{\frac{5}{\lambda_3},\frac{4}{5}} r_{\frac{3}{\lambda_2},2} r_{\frac{2}{\lambda_1},\frac{7}{2}}}{\lambda_1 \lambda_2 \lambda_3 \psi^7(t)}.
\end{align*}
}
It can be seen that the upper bound of \( x_1 \)  demonstrates uniform hyperexponential convergence in initial conditions.
Given that \( x_1 \), \( \sg_2 \), and \( \sg_3 \) converge uniformly to zero, the behavior of \( x_2 \) can be inferred from \( \sg_2 \) as follows:
\crb{\begin{gather*}
|x_2(t)| \le C_1(t)\,|x_1(0)| + C_2(t)\,|\sigma_2(0)| + C_3(t)\,|\sigma_3(0)| \\
+ D_1(t)\,\|d_1\|_\infty + D_2(t)\,\|d_2\|_\infty + D_3(t)\,\|d_3\|_\infty,
\end{gather*}
where
\begin{align*}
C_1(t) &= \lambda_1 \psi(t)\, \E^{-\lambda_1 \frac{t}{2}(t+2)}, \\[2mm]
C_2(t) &= \E^{-\lambda_2 t \left(\frac{t^2}{3} + t + 1\right)} + \lambda_1 \psi(t)\, \frac{3 \E^{-\lambda_1 \frac{t}{2}(t+2)}}{2 \lambda_2 - 3 \lambda_1}, \\[1mm]
C_3(t) &= \frac{5 \E^{-\lambda_2 t \left(\frac{t^2}{3} + t + 1\right)}}{3 \lambda_3 - 5 \lambda_2} 
+ \frac{15 \lambda_1 \psi(t) \E^{-\lambda_1 \frac{t}{2}(t+2)}}{(3 \lambda_3 - 5 \lambda_2)(2 \lambda_2 - 3 \lambda_1)}, \\[1mm]
D_1(t) &= \frac{\lambda_1}{\lambda_3 \lambda_2} \Bigg(
\frac{\lambda_2 r_{\frac{5}{\lambda_3},\frac{1}{5}} r_{\frac{3}{\lambda_2},1}}{\psi^3(t)} 
+ \frac{r_{\frac{5}{\lambda_3},\frac{4}{5}} r_{\frac{3}{\lambda_2},2}}{\psi^6(t)}
\\& + \frac{\lambda_3 r_{\frac{3}{\lambda_2},\frac{1}{3}}}{\psi(t)}
+ \frac{\lambda_2 r_{\frac{5}{\lambda_3},\frac{1}{5}} r_{\frac{3}{\lambda_2},1} r_{\frac{2}{\lambda_1},2}}{\psi^3(t)} \\
&\quad + \frac{r_{\frac{5}{\lambda_3},\frac{4}{5}} r_{\frac{3}{\lambda_2},2} r_{\frac{2}{\lambda_1},\frac{7}{2}}}{\psi^6(t)}
+ \lambda_3 r_{\frac{3}{\lambda_2},\frac{1}{3}} r_{\frac{2}{\lambda_1},\frac{1}{2}}
+ \frac{\lambda_2 \lambda_3 r_{\frac{2}{\lambda_1},\frac{1}{2}}}{\lambda_1} \Bigg), \\[1mm]
D_2(t) &= \frac{1}{\lambda_2 \lambda_3} \Bigg(
\frac{\lambda_1 r_{\frac{5}{\lambda_3},\frac{3}{5}} r_{\frac{3}{\lambda_2},\frac{5}{3}}}{\psi^5(t)}
+ \frac{\lambda_2 r_{\frac{5}{\lambda_3},\frac{2}{5}} r_{\frac{3}{\lambda_2},\frac{4}{3}}}{\psi^4(t)}
\\ &+ \frac{\lambda_3 r_{\frac{3}{\lambda_2},\frac{2}{3}}}{\psi^2(t)}
+ \frac{\lambda_1 r_{\frac{5}{\lambda_3},\frac{3}{5}} r_{\frac{3}{\lambda_2},\frac{5}{3}} r_{\frac{2}{\lambda_1},3}}{\psi^5(t)} \\
&\quad + \frac{\lambda_2 r_{\frac{5}{\lambda_3},\frac{2}{5}} r_{\frac{3}{\lambda_2},\frac{4}{3}} r_{\frac{2}{\lambda_1},\frac{5}{2}}}{\psi^4(t)}
+ \frac{\lambda_3 r_{\frac{3}{\lambda_2},\frac{2}{3}} r_{\frac{2}{\lambda_1},\frac{3}{2}}}{\psi^2(t)} \Bigg), \\[1mm]
D_3(t) &= \frac{r_{\frac{5}{\lambda_3},\frac{4}{5}} r_{\frac{3}{\lambda_2},2}}{\lambda_2 \lambda_3 \psi^6(t)} \left(1 + r_{\frac{2}{\lambda_1},\frac{7}{2}}\right),
\end{align*}
}
which implies boundedness of this variable, with the asymptotic bound proportional to $ d_1$ with the gain $ \crb{\lam_3}r_{\frac{2}{\lam_1},\frac{1}{2}}(1+\lam_1\lam_2^{-1}r_{\frac{3}{\lam_2},\frac{1}{3}})$, and hyperexponential decay of vanishing \crb{in} initial conditions.

From the expression for \( \sg_3 \), we have the inequality for \( x_3(t) \):

 \[
|x_3(t)| \leq |\sg_3(t)| + \lam_1 \psi(t) |x_2(t)| + \lam_1 |x_1| + \lam_2 \psi^2(t) |\sg_2(t)| = {\cb \sum_{i=1}^4 T_i(t),}
\]
{\cb 
where 
\[
T_1\crb{(t)} = |\sigma_3(t)| \leq \E^{-\lambda_3 t P_3(t)} |\sigma_3(0)| + \mathcal{O}\left(\psi^{-1}(t) + \psi^{-4}(t)\right)\|d\|_\infty,
\]
with $P_3(t) = \frac{t^4}{5} + t^3 + 2t^2 + 2t + 1$ and $\mathcal{O}$ is constant term;
\[
T_2\crb{(t)} = \lambda_1 \psi(t) |x_2(t)| 
\leq \lambda_1 \psi(t) \Big(|\sigma_2(t)| + \lambda_1 \psi(t) |x_1(t)|\Big),
\]
where:
\begin{gather*}
|\sigma_2(t)| \leq \E^{-\lambda_2 t P_2(t)} |\sigma_2(0)| + \mathcal{O}\left(\psi^{-3}(t)\right)\|d\|_\infty, \\
|x_1(t)| \leq \E^{-\lambda_1 t P_1(t)} |x_1(0)| + \mathcal{O}\left(\psi^{-1}(t)\right)\|d\|_\infty,
\end{gather*}
and $P_2(t) = \frac{t^2}{3} + t + 1$, $P_1(t) = \frac{t(t+2)}{2}$;
\[
T_3\crb{(t)} = \lambda_1 |x_1(t)| \leq \lambda_1 \E^{-\lambda_1 t P_1(t)} |x_1(0)| + d_0,
\]
where $d_0>0$ represents a constant;
\begin{gather*}
T_4\crb{(t)} = \lambda_2 \psi^2(t) |\sigma_2(t)| \\ \leq \lambda_2 \psi^2(t) \left(\E^{-\lambda_2 t P_2(t)} |\sigma_2(0)| + \mathcal{O}\left(\psi^{-3}(t)\right)\|d\|_\infty\right).\end{gather*}

The terms $T_1$ and $T_4$ are bounded and exhibit uniform convergence. The  term $T_2$ grows  due to  $x_2$ is bounded and the unbounded growth of \( \psi(t) \). 
\crb{Under the conditions of Corollary~\ref{cor:1}, where  \(\psi(t)\) is saturated by replacing it in the control algorithm by $\min\{\psi(t),\Psi_{\epsilon,\Delta}\}$,  such that we can use the following upper bound in the computations above:}
\[
\psi(t) \le \Psi_{\epsilon,\Delta}, \quad \forall t \ge 0,
\]
we then have
\begin{equation}
|x_3(t)| \leq \underbrace{\beta(|x(0)|, t)}_{\mathcal{KL}\text{-term}} + \underbrace{\left(\gamma_1 + \lambda_1 \Psi_{\epsilon,\Delta} \gamma_2 + \lambda_2 \Psi_{\epsilon,\Delta}^2 \gamma_3\right)\|d\|_\infty}_{\text{ISS gain}},
\end{equation}
where
\(
\gamma_1,
\gamma_2, 
\gamma_3
\) are \crb{some} positive constants. 
From the estimates derived for $x_1$ and $x_2$ we can conclude that for any limit on the amplitude of perturbations and the upper bound on $\psi$, the component $x_1$ will globally converge to a neighbourhood of the origin with hyperexponential decay in initial conditions. Once $x_1$ being in a given neighbourhood, and the growth of $\psi$ is canceled, the variables $x_2$ and $x_3$ will be bounded and converging to a domain proportional to the amplitude of disturbances. The whole closed-loop system demonstrates ISS behavior with a uniform bound in $x_1$. This completes the proof.}
\end{proof}
{\cb \begin{rem}
 For \( i \geq 2 \), the use of a saturation function with \( \psi \) ensures that the input control remains bounded. \hfill $\triangle$
\end{rem}
In the following section, let us analyze the effects of discretization on the convergence of the state and the robustness of the control strategy.
}

\section{Discretization of the controlled system~\eqref{eq:sysm-0}}
The main result presented in the previous section {\cb deals with} a linear time-varying  {\cb system~\eqref{eq:sysm-0}--\eqref{eq:u} } with an external input $d(t)$. Due to the strictly increasing nature of the time-varying control gain $\psi(t)$, the explicit Euler discretization becomes unsuitable, as the unbounded growth of $\psi(t)$ can lead to numerical instability. In contrast, the implicit Euler discretization offers a robust alternative and can be effectively applied to this system \cite{Brogliato2021, Butcher2008}.

 {\cb It is worth noting that in \cite{wang2024hyperexponential}, a similar problem has been considered, including the direct application of the implicit Euler discretization method. This approach simplifies the analysis and ensures numerical stability when dealing with rapidly growing control gains. In our case, however, the control input and the dynamic system structure depend on  $\omega(x)$ and $\Omega(x,\psi)$. These additional terms introduce complexities that must be carefully addressed to ensure the accurate application of the implicit Euler method to the resulting system dynamics (they are nonlinearly depending on $x$, then it is difficult to resolve the discretized dynamics with respect to future state value). To effectively apply this discretization method,  we propose transforming the system equations into $\sigma$-coordinates, applying the implicit discretization in this space, and subsequently transforming the results back to the original $x$-coordinates.}





The transformation of the system equations into \(\sigma\)-coordinates has the following form:
\[
\sigma(t) = 
\begin{pmatrix}
\sigma_1(t) \\ 
\sigma_2(t) \\ 
\vdots \\
\sigma_n(t)
\end{pmatrix} =  
S(t) x(t),
\]
where 
\[
S(t) = 
\begin{bmatrix}
1      & 0      & \cdots & 0      & 0      \\
P_{12}^{\beta_2}(\psi)      & 1      & \cdots & 0      & 0      \\
\vdots & P_{13}^{\alpha_3}(\psi) & \ddots & \vdots & \vdots \\
P_{1n-1}^{\beta_{n-1}}(\psi)      & \cdots      & \cdots & 1      & 0      \\
P_{1n}^{\beta_n}(\psi)      & P_{2n}^{\alpha_{2n-2}}(\psi) & \cdots & P_{(n-1)n}^{\gamma}(\psi)      & 1
\end{bmatrix}.
\]
and \( P_{ji}^{\beta_i}(\psi), P_{ji}^{\alpha_i}(\psi), \ldots, P_{ji}^{\gamma}(\psi) \) {\cb  for} \(i,j = 1, \ldots, n\), are polynomials of \(\psi(t)\), and their respective degrees are given as \(\beta_i\), \(\alpha_i\), and \(\gamma\). These degrees satisfy the following conditions:
\[
\beta_l < \beta_{l+1}, \ \cdots, \ \alpha_l < \alpha_{l+1}, \ \text{and} \quad \beta_n > \alpha_{n-2} >\cdots > \gamma.
\]

The inverse transformation back to \(x(t)\)-coordinates is then given by:
\[
x(t) = S(t)^{-1} \sigma(t).
\]
Note that $\det(S) = 1$, and  $S^{-1}(t)$ is well defined and it has the form \begin{gather*}
S^{-1}(t) = 
\begin{bmatrix}
1 & 0 & 0 & \cdots & 0 \\
-P_{12}^{\beta_2}(\psi) & 1 & 0 & \cdots & 0 \\
P_{12}^{\beta_2}(\psi)P_{13}^{\alpha_3}(\psi) - P_{23}^{\beta_3}(\psi) & -P_{33}^{\alpha_3}(\psi) & 1 & \cdots & 0 \\
\vdots & \vdots & \vdots & \ddots & 0 \\
\star & \star & \star & \cdots & 1
\end{bmatrix}.
\end{gather*}

Consider the dynamics of $\sigma$ \crb{under the proposed control law for $m=n$}:
\beq\label{eq:sigma_dis}
\dot{\sigma}(t) = M(t) \sigma(t) + L(t) d(t),
\eeq

\[
M(t) = \begin{bmatrix}
-\psi(t) & 1 & 0 & \cdots & 0 \\
0 & -\psi(t)^2 & 1 & \cdots & 0 \\
0 & 0 & -\psi(t)^3 & \ddots & \vdots \\
\vdots & \vdots & \vdots & \ddots & 1 \\
0 & 0 & 0 & \cdots & -\psi(t)^n
\end{bmatrix},\]
and
\[ L(t) =   
\begin{bmatrix}
1      & 0      & \cdots & 0      & 0      \\
P_{12}^{\mu_2}(\psi)      & 1      & \cdots & 0      & 0      \\
\vdots & P_{23}^{\eta_3}(\psi) & \ddots & \vdots & \vdots \\
P_{1n-1}^{\mu_{n-1}}(\psi)      & \cdots      & \cdots & 1      & 0      \\
P_{1n}^{\mu_n}(\psi)      & P_{2n}^{\eta_{n-2}}(\psi) & \cdots & P_{(n-1)n}^{\nu}(\psi)      & 1
\end{bmatrix},
\]
where \( P_{ji}^{\mu_i}(\psi), P_{ji}^{\eta_i}(\psi), \ldots, P_{ji}^{\nu}(\psi) \) {\cb  for}  \(i,j = 1, \ldots, n\), are polynomials of \(\psi(t)\), and their respective degrees are given as \(\mu_i\), \(\eta_i\), and \(\nu\). These degrees satisfy the following conditions:
\[
\mu_l < \mu_{l+1}, \cdots, \quad \eta_l < \eta_{l+1},  \quad \text{and} \quad \mu_n > \eta_{n-2} > \cdots> \nu.
\]

Let \( h > 0 \) denote a constant discretization step, and define the discretization time instants by \( t_k = hk \), where \( k \in \mathbb{Z}^+ \). Applying the implicit Euler discretization method to the system equation (\ref{eq:sigma_dis}) results in the following update for \( k \in \mathbb{Z}^+ \):

\[
\zeta_{k+1} = Z(t_{k+1}) \left( \zeta_k + h L(t_{k+1}) d_{k+1} \right), 
\]
where \( \zeta_k \in \mathbb{R}^n \) is an approximation of \( \sigma_k = \sigma(t_k) \), meaning that \( \zeta_k \to \sigma_k \) as \( h \to 0 \) {\cb(on any compact interval of discretization time \cite{Butcher2008})}\crb{, and} sequences \( d_k = d(t_k) \) represent the disturbance at the discretization time \( t_k \).
The \crb{matrix} \( Z(t) \) is defined as:
\begin{gather*}
Z(\rho) = \begin{bmatrix}I_n - hM(t)\end{bmatrix}^{-1}
\implies \\ 
Z(\rho) = 
\begin{bmatrix}
\frac{1}{\rho_1} & \frac{h}{\rho_1 \rho_2} & \frac{h^2}{\rho_1 \rho_2 \rho_3} & \cdots &  \frac{h^{n-1}}{\rho_1 \rho_2 \cdots \rho_n} \\
0 & \frac{1}{\rho_2} & \frac{h}{\rho_2 \rho_3} & \cdots &  \frac{h^{n-2}}{\rho_2 \rho_3 \cdots \rho_n} \\
0 & 0 & \frac{1}{\rho_3} & \cdots &  \frac{h^{n-3}}{\rho_3 \rho_4 \cdots \rho_n} \\
\vdots & \vdots & \vdots & \ddots & \vdots \\
0 & 0 & 0 & \cdots & \frac{1}{\rho_n}
\end{bmatrix},
\end{gather*}
\[
\rho_i = 1 + h\psi(t)^i, \quad i = 1, 2, \dots, n.
\]
Using \( L(t) \), we have:
\begin{gather*} Z(\rho) \times L(t) = 
\begin{bmatrix}
\frac{1}{\rho_1} & \frac{h}{\rho_1 \rho_2} & \frac{h^2}{\rho_1 \rho_2 \rho_3} & \cdots &  \frac{h^{n-1}}{\rho_1 \rho_2 \cdots \rho_n} \\
0 & \frac{1}{\rho_2} & \frac{h}{\rho_2 \rho_3} & \cdots &  \frac{h^{n-2}}{\rho_2 \rho_3 \cdots \rho_n} \\
0 & 0 & \frac{1}{\rho_3} & \cdots &  \frac{h^{n-3}}{\rho_3 \rho_4 \cdots \rho_n} \\
\vdots & \vdots & \vdots & \ddots & \vdots \\
0 & 0 & 0 & \cdots & \frac{1}{\rho_n}
\end{bmatrix},
\end{gather*}
which allows us to deduce that \( \lim_{t \to \infty} \left( Z(t) L(t) \right) = 0 \).

\crb{In this section}, we assume that the sequence \( d_k \) is bounded, i.e., \( d_k \in \ell_n^\infty \).

Now, returning to the coordinate \( x \), the closed-loop system is given by the following Implicit Euler discretization:
\beq\label{eq:x_dis}
\xi_{k+1} = S(t_{k+1})^{-1}  Z(t_{k+1}) \left( S(t_{k+1}) \xi_k + h L(t_{k+1}) d_{k+1} \right),
\eeq
where \( \xi_k \in \mathbb{R}^n \) is an approximation of \( x_k = x(t_k) \). {\cb The relationship \(\xi_k \to x_k\) as \(h \to 0\) on any bounded interval of discretization time was established in both 
Theorem~213A and Theorem~213B of \cite{Butcher2008}.}
\begin{thm}\label{theom:dis}
Let the auxiliary variable \( \sigma \) be chosen such that 
$
\beta_l < \beta_{l+1}, \quad \cdots, \quad \alpha_l < \alpha_{l+1}, \quad \text{and} \quad \beta_n > \alpha_{n-2} > \cdots > \gamma, \quad \mu_l < \mu_{l+1}, \quad \cdots, \quad \eta_l < \eta_{l+1}, \quad \text{and} \quad \mu_n > \eta_{n-2} > \cdots > \nu.
$
Then, the system \eqref{eq:x_dis} {\cb is} ISS. Moreover, \crb{the origin is uniformly hyperexponentially stable} {\cb if $|d|_\infty=0$}.
\end{thm}
\begin{proof}
To analyze the asymptotic system's properties, such as hyperexponential stability and robustness, we examine the following limits:
\[
\lim_{t \to \infty} \left( S(t)^{-1} Z(t) S(t) \right) \quad \text{and} \quad \lim_{t \to \infty} \left( S(t)^{-1} Z(t) L(t) \right).
\]
Moreover, we have the following expression for the first limit:
\begin{gather*}
\lim_{t \to \infty} \left( S(t)^{-1} Z(t) S(t) \right) 
\crb{
= -\frac{1}{h^{\,n-1}}
\begin{bmatrix}
0 & 0 & 0 & \cdots & 0 \\[6pt]
h^{\,n-2} & 0 & 0 & \cdots & 0 \\[6pt]
h^{\,n-3} & h^{\,n-2} & 0 & \cdots & 0 \\[6pt]
\vdots & \vdots & \ddots & \ddots & \vdots \\[6pt]
1 & h & \cdots & h^{\,n-2} & 0
\end{bmatrix},}
\end{gather*}
which implies the hyperexponential stability of the system \eqref{eq:x_dis}. Furthermore, the second limit is the same:
\begin{gather*}
\lim_{t \to \infty} \left( S(t)^{-1} Z(t) S(t) \right) \crb{
= -\frac{1}{h^{\,n-1}}
\begin{bmatrix}
0 & 0 & 0 & \cdots & 0 \\[6pt]
h^{\,n-2} & 0 & 0 & \cdots & 0 \\[6pt]
h^{\,n-3} & h^{\,n-2} & 0 & \cdots & 0 \\[6pt]
\vdots & \vdots & \ddots & \ddots & \vdots \\[6pt]
1 & h & \cdots & h^{\,n-2} & 0
\end{bmatrix},}
\end{gather*}
which implies that the system is ISS with respect to disturbances \( d \), which completes the proof.
\end{proof}
\section{Illustrative examples}
{\cb In this section we give the explicit computations for several cases.}
 Let us consider a chain of integrators for \( n = 3 \).
We define the following auxiliary variables to facilitate the analysis of the system dynamics:
\[
\begin{aligned}
\sigma_1(t) &= x_1(t), \\
\sigma_2(t) &= x_2(t) + \psi(t) x_1(t), \\
\sigma_3(t) &= x_3(t) + \psi(t) x_2(t) + x_1(t) + \psi(t)^2 \sigma_2(t),
\end{aligned}
\]
where \( \psi(t) = 1 + t \).

The control law is constructed as:
\beq\label{eq:u3_dis}
\bag
u(t) &= -\psi(t)^3 \sigma_3(t) - \psi(t) \left( x_3(t) + 2 \sigma_2(t) \right) - 2 x_2(t)  \\ & \quad 
 + \psi(t)^2 \left( \psi(t)^2 \sigma_2(t) - \sigma_3(t) \right).
 \eag
\eeq
\begin{cor} 
Under the conditions of Theorem \ref{theom:dis}, the system \eqref{eq:x_dis} for $n=3$ \crb{is ISS}. Furthermore, the system \crb{is uniformly hyperexponentially stable at the origin if $|d|_\infty=0$}. 
\end{cor}
\begin{proof}
First, we examine the dynamics of the auxiliary variables, which evolve according to the following differential equations:
\[
\begin{aligned}
\dot \sigma_1(t) &= -\psi(t) \sigma_1(t) + d_1(t) + \sigma_2(t), \\
\dot \sigma_2(t) &= -\psi(t)^2 \sigma_2(t) + d_2(t) + \psi(t) d_1(t) + \sigma_3(t), \\
\dot \sigma_3(t) &= -\psi(t)^3 \sigma_3(t) + \psi(t)(1 + \psi(t)) d_2(t)\\ &\quad  + (1 + \psi(t)^3) d_1(t) + d_3(t).
\end{aligned}
\]
Next, we define the coordinate transformation:

\[
\sigma(t) = 
\begin{pmatrix}
\sigma_1(t) \\ 
\sigma_2(t) \\ 
\sigma_3(t)
\end{pmatrix} 
= 
S(t) x(t),
\]
\[
S(t) = 
\begin{bmatrix}
1 & 0 & 0 \\ 
P_{12}^1(\psi) & 1 & 0 \\ 
P_{13}^3(\psi) & P_{13}^2(\psi) & 1
\end{bmatrix},
\]
where: $
P_{12}^1(\psi) = \psi(t), \;
P_{13}^3(\psi) = \psi(t)^3 + 1, \;
P_{23}^2(\psi) = (\psi(t) + 1)\psi(t).
$
The inverse matrix \( S(t)^{-1} \) is:
\[
S(t)^{-1} = 
\begin{bmatrix}
1 & 0 & 0 \\ 
P_{12}^1(\psi) & 1 & 0 \\ 
P_{13}^2(\psi)  & P_{23}^2(\psi) & 1
\end{bmatrix},
\]
where:
$
P_{12}^1(\psi) = -\psi(t), \;
P_{13}^2(\psi) = \psi(t)^2 - 1, P_{23}^2(\psi) = -\psi(t)^2 -\psi(t).
$

Since \( \operatorname{det}(S(t)) = 1 \), the transformation is invertible, preserving the structure of the system.

By applying the chain rule, we derive the dynamics of the transformed variables \( \sigma(t) \):
\[
\dot{\sigma}(t) = M(t) \sigma(t) + L(t) d(t),
\]
where \( M(t) \) and \( L(t) \) are time-varying matrices defined as:

\begin{gather*}
M(t) = \begin{bmatrix}
-\psi(t) & 1 & 0 \\
0 & -\psi(t)^2 & 1 \\
0 & 0 & -\psi(t)^3
\end{bmatrix}, \\ 
L(t) = \begin{bmatrix}
1 & 0 & 0 \\
P_{12}^{1}(\psi) & 1 & 0 \\
P_{13}^{3}(\psi)  & P_{23}^2(\psi) & 1
\end{bmatrix},
\end{gather*}
where $P_{12}^{1}(\psi)=\psi(t)$, $P_{13}^{3} = \psi(t)^3 + 1$, and $P_{23}^2(\psi)=(\psi(t) + 1)\psi(t)$.

To discretize the system, we {\cb use} the following implicit {\cb Euler} discretization scheme:

\[
\zeta_{k+1} = \zeta_k + h \left( M(t_{k+1}) \zeta_{k+1} + L(t_{k+1}) d_{k+1} \right),
\]
where \( \zeta_k \in \mathbb{R}^3 \) is an approximation of \( \sigma_k = \sigma(t_k) \).

Rearranging this, we obtain:

\[
\zeta_{k+1} = Z(t_{k+1}) \left( \zeta_k + h L(t_{k+1}) d_{k+1} \right),
\]
where \( Z(t) \) is a matrix that depends on \( \psi(t) \) and is given by:
\[
Z(t) = 
\begin{bmatrix}
\frac{1}{\rho_1} & \frac{h}{\rho_1 \rho_2} & \frac{h^2}{\rho_1 \rho_2 \rho_3} \\
0 & \frac{1}{\rho_2} & \frac{h}{\rho_2 \rho_3} \\
0 & 0 & \frac{1}{\rho_3}
\end{bmatrix},
\]
where 
\[
\rho_i = 1 + h \psi(t)^i, \quad i = 1, 2, 3.
\]

Finally, we describe the closed-loop system, which is given by:
\begin{align}\label{eq:sys_3}
\xi_{k+1} = S(t_{k+1})^{-1} \left[ Z(t_{k+1}) \left( S(t_{k+1}) \xi_k + h L(t_{k+1}) d_{k+1} \right) \right],
\end{align}
where \( \xi_k \in \mathbb{R}^3 \) is an approximation of \( x_k = x(t_k) \).
As \( t \to \infty \), the behavior of the system can be characterized by the following limits:

\[
\lim_{t \to \infty} \left( S(t)^{-1} Z(t) S(t) \right) = -\frac{1}{h^2} \begin{bmatrix} 0 & 0 & 0 \\ h & 0 & 0 \\ 1 & h & 0 \end{bmatrix},
\]
which \crb{is a nilpotent matrix}. 
For robustness, consider the following term: 
\[
\lim_{t \to \infty} \left( S(t)^{-1} Z(t) L(t) \right) = -\frac{1}{h^2} \begin{bmatrix} 0 & 0 & 0 \\ h & 0 & 0 \\ 1 & h & 0 \end{bmatrix}.
\]
Thus, we conclude that the system \crb{is} robust stability and \crb{has a} faster than \crb{any}  exponential convergence to the origin.

\end{proof}

Let us consider a chain of integrators for \( n = 4 \).
We define the following auxiliary variables:
 \[
\begin{aligned}
\sigma_1(t) &= x_1(t), \\
\sigma_2(t) &= x_2(t) + \psi(t) x_1(t), \\
\sigma_3(t) &= x_3(t) + \psi(t) x_2(t) + x_1(t) + \psi(t)^2 \sigma_2(t),\\
\sigma_4(t) &= x_4(t) +x_3(t)(1+\psi(t))\psi(t)  +x_2(t)(2+\psi(t)^3) \\
& \qquad + \psi(t)^2x_1(t) + 2\psi(t) \sigma_2(t) +\psi(t)^3\sg_3(t).
\end{aligned}
\]
The control law is constructed as:
\beq\label{eq:u4_dis} 
\begin{aligned}
u(t) = &- \psi(t)^4 \sigma_4(t) - x_4(t) \left( \psi(t) + \psi(t)^2 + \psi(t)^3 \right)\\ & \quad - x_3(t) \left( 3 + 4\psi(t) + \psi(t)^3 + \psi(t)^4 + \psi(t)^5 \right) \\
&- x_2(t) \left( \psi(t)^2 + 2\psi(t)^3 + \psi(t)^6 \right)- 3 \psi(t)^2 \sigma_3 \\ &\quad 
- x_1(t) \left( 4\psi(t) + \psi(t)^5 \right)
- 2 \sigma_2 (1 + \psi(t)^4).
\end{aligned}
\eeq
\begin{cor}
Under the \crb{conditions of} Theorem \ref{theom:dis}, the system \eqref{eq:x_dis} for $n=4$ \crb{is} ISS. Furthermore, the system  \crb{is uniformly hyperexponentially stable at the origin} if $|d|_\infty=0$.
\end{cor}
\begin{proof}
We have:
 \[
\begin{aligned}
\dot \sigma_1(t) &= -\psi(t)\sg_1(t) +d_1(t) +\sg_2(t), \\
\dot \sigma_2(t) &= -\psi(t)^2\sg_2(t) + d_2(t) +\psi(t)d_1(t) +\sg_3(t), \\
\dot \sigma_3(t) &= -\psi(t)^3\sg_3(t) + \psi(t)(1+\psi(t))d_2(t)\\  & \quad 
 +(1+\psi(t)^3)d_1(t) +d_3(t),\\
\dot \sg_4(t) &= -\psi(t)^4\sg_4(t) + d_1(t)\left(3\psi(t)^2+\psi(t)^3+\psi(t)^6\right)
\\  & \quad  + d_2(t)\left(2+2\psi(t)+\psi(t)^3+\psi(t)^4+\psi(t)^5\right) \\
&\qquad + d_3(t)\left(\psi(t)+\psi(t)^2+\psi(t)^3\right)+d_4(t),
\end{aligned}
\]
Transformation of coordinates is defined as:

\begin{gather*}
\sigma(t) 
= 
\begin{bmatrix}
1 & 0 & 0 & 0 \\ 
P_{12}^1(\psi) & 1 & 0 & 0 \\ 
P_{13}^3(\psi) & P_{23}^2(\psi) & 1 & 0 \\ 
P_{14}^6(\psi) & P_{24}^5(\psi) & P_{34}^4(\psi) & 1
\end{bmatrix}x(t)= S(t)  x(t),
\end{gather*}
where 
$
P_{12}^1(\psi) = \psi(t), \; 
P_{13}^3(\psi) = \psi(t)^3 + 1, \;
P_{23}^2(\psi) = (\psi(t) + 1)\psi(t),
P_{14}^6(\psi) = \psi(t)^6 + \psi(t)^3 + 3\psi(t)^2, \;
P_{24}^5(\psi) = 2 + 2\psi(t) + \psi(t)^3 + \psi(t)^4 + \psi(t)^5, \;
P_{34}^4(\psi) = \psi(t)(1 + \psi(t) + \psi(t)^2).
$

{\cb Consequently,}
\[ 
x(t) = S(t)^{-1}  \sigma(t) = 
\begin{bmatrix}
1 & 0 & 0 & 0 \\
P_{12}^1(\psi) & 1 & 0 & 0 \\
P_{13}^2(\psi) & P_{23}^2(\psi) & 1 & 0 \\
P_{14}^3(\psi) & P_{24}^4(\psi) & P_{34}^3(\psi)  & 1
\end{bmatrix}
\sigma(t),\]
where 
$
P_{12}^1(\psi) = -\psi(t), \;
P_{13}^2(\psi) = \psi(t)^2 - 1, \;P_{23}^2(\psi) = -\psi(t)(\psi(t) +1), \;
P_{14}^3(\psi) = 3\psi(t) - \psi(t)^3, \;
P_4^4(\psi) = \psi(t)^2 - 2\psi(t) + \psi(t)^3 + \psi(t)^4 - 2, \;
P_{34}^3(\psi) = -\psi(t)(\psi(t) + \psi(t)^2 + 1)
$, with  $\operatorname{det}(S(t)) = 1$, 
and then
\[
\dot{\sigma}(t) = M(t) \sigma(t) + L(t) d(t),
\]
\[
M(t) = \begin{bmatrix}
-\psi(t) & 1 & 0 &0\\
0 & -\psi(t)^2 & 1&0 \\
0 & 0 & -\psi(t)^3&1
 \\
0 & 0 & 0& -\psi(t)^4
\end{bmatrix}, \]
\[
L(t) = \begin{bmatrix}
1 & 0 & 0  & 0 \\
P_{12}^1(\psi) & 1 & 0  & 0 \\
P_{13}^3(\psi) & P_{23}^2(\psi) & 1 & 0 \\
P_{14}^6(\psi) & P_{24}^5(\psi) & P_{34}^4(\psi) & 1
\end{bmatrix},
\]
where:
$
P_{12}^1(\psi) = \psi(t), \;
P_{13}^3(\psi) = \psi(t)^3 + 1, \;
P_{23}^2(\psi) = (\psi(t) + 1) \psi(t),
P_{14}^6(\psi) = 3\psi(t)^2 + \psi(t)^3 + \psi(t)^6, \;
P_{24}^5(\psi) = 2 + 2\psi(t) + \psi(t)^3 + \psi(t)^4 + \psi(t)^5, \;
P_{34}^4(\psi) = \psi(t) + \psi(t)^2 + \psi(t)^3.
$
{\cb Therefore, applied  implicit discretization takes the form:}
\[\zeta_{k+1} = \zeta_k + hM(t_{k+1})\zeta_{k+1} + L(t_{k+1}) d_{k+1}
\]
where \( \zeta_k \in \mathbb{R}^4 \) is an approximation of \( \sigma_k = \sigma(t_k) \).
Then,
\[\zeta_{k+1} = Z(t_{k+1})\Bigg(\zeta_k + h L(t_{t+1})d_{k+1}\Bigg),\]
with 
\[
Z(t) = 
\begin{bmatrix} 
\frac{1}{\rho_1} & \frac{h}{\rho_1 \rho_2} & \frac{h^2}{\rho_1 \rho_2 \rho_3} & \frac{h^3}{\rho_1 \rho_2 \rho_3 \rho_4} \\
0 & \frac{1}{\rho_2} & \frac{h}{\rho_2 \rho_3} & \frac{h^2}{\rho_2 \rho_3 \rho_4} \\
0 & 0 & \frac{1}{\rho_3} & \frac{h}{\rho_3 \rho_4} \\
0 & 0 & 0 & \frac{1}{\rho_4}
\end{bmatrix},
\]
where 
\[
\rho_i = 1 + h \psi(t)^i, \quad i = 1, 2, 3, 4.
\]
The closed-loop system is given by:
\[
\xi_{k+1} = S(t_{k+1})^{-1} \left[ Z(t_{k+1}) \left( S(t_{k+1}) \xi_k + h L(t_{k+1}) d_{k+1} \right) \right]
\]
where \( \xi_k \in \mathbb{R}^4 \) is an approximation of \( x_k = x(t_k) \).
We have:
\[
\lim_{t \to \infty} \left( S(t)^{-1} Z(t) S(t) \right) =  
-\frac{1}{h^3} \begin{bmatrix}
0 & 0 & 0 & 0 \\
h^2 & 0 & 0 & 0 \\
h & h^2 & 0 & 0 \\
1 & h & h^2 & 0
\end{bmatrix},
\]
For robustness, we now check the following limit: 
\[
\lim_{t \to \infty} \left( S(t)^{-1} Z(t) L(t) \right) = -\frac{1}{h^3} \begin{bmatrix}
0 & 0 & 0 & 0 \\
h^2 & 0 & 0 & 0 \\
h & h^2 & 0 & 0 \\
1 & h & h^2 & 0
\end{bmatrix}.
\]
Thus, we conclude that the system \crb{is} robust stability and \crb{has a} faster-than-exponential convergence to the origin.
\end{proof}


\section{Simulations}
\begin{figure}[ht!]
    \centering
    \includegraphics[width=0.5\textwidth]{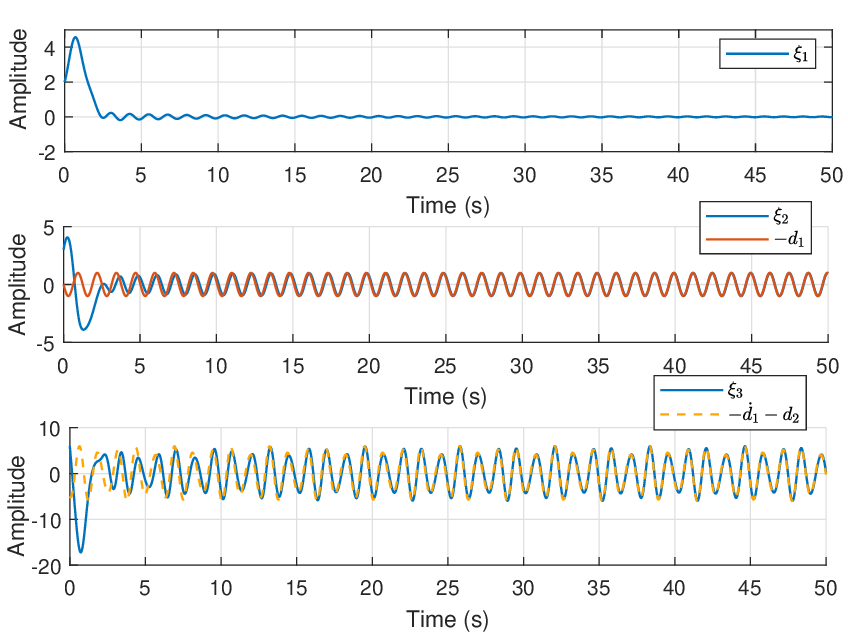}
    \caption{Objective vs time.}
    \label{fig:fig1}
\end{figure}

\begin{figure}[ht!]
    \centering
    \includegraphics[width=0.5\textwidth]{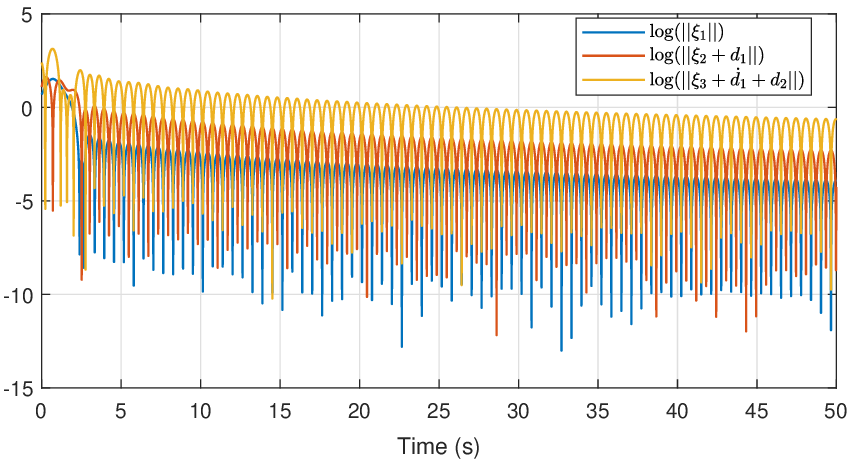}
    \caption{Error norms in
logarithmic scale vs time.}
    \label{fig:fig2}
\end{figure}

Let us illustrate the performance of the proposed hyperexponential stabilizer for the system described by \eqref{eq:sys_3} through numerical simulations. For \( h = 0.001 \), the simulation results are presented in Figs. \ref{fig:fig1} and \ref{fig:fig2}. The perturbations  are {\cb considered} as follows:
\(
d_1(t) = \sin(5t), \quad
d_2(t) = \sin(7t), \quad
d_3(t) = \cos(3t) - \rnd(1).
\)
The top plot shows the state trajectory of \( \xi_1 \), which converges to zero over time. The middle plot illustrates that the state \( \xi_2 \) converges to \( -d_1 \), while the bottom plot demonstrates that the state \( \xi_3 \) converges to \( -d_2 - \dot{d}_1 \). Fig. \ref{fig:fig2} displays the boundedness of the norms, plotted on a logarithmic scale.
\section{Conclusions}
A new control strategy for a chain of integrators subject to unmatched perturbations is proposed. In its continuous\crb{-time} form, it ensures hyperexponential convergence for the state variable $x_1$, while $x_2$ remains bounded. For the other states, ISS is achieved by saturating the growing gain. A simple discrete-time implementation is also provided, guaranteeing uniform hyperexponential convergence of state trajectories to the origin in the \crb{absence} of {\cb both} matched and unmatched inputs. The main results are illustrated through  numerical simulations.
%
%
\bibliographystyle{elsarticle-num}
\bibliography{references}

\begin{thebibliography}{10}
\expandafter\ifx\csname url\endcsname\relax
  \def\url#1{\texttt{#1}}\fi
\expandafter\ifx\csname urlprefix\endcsname\relax\def\urlprefix{URL }\fi
\expandafter\ifx\csname href\endcsname\relax
  \def\href#1#2{#2} \def\path#1{#1}\fi

\bibitem{Chernousko2008}
F.~L. Chernous'ko, I.~M. Ananievski, S.~A. Reshmin, Control of nonlinear
  dynamical systems, Springer, 2008.

\bibitem{Isidori1995}
A.~Isidori, Nonlinear Control Systems, 3rd Edition, Communications and Control
  Engineering Series, Springer-Verlag, 1995.

\bibitem{Khalil2002}
H.~K. Khalil, Nonlinear Systems, 3rd Edition, Prentice Hall, 2002.

\bibitem{Utkin1992}
V.~I. Utkin, Sliding Modes in Control Optimization, Springer-Verlag, 1992.

\bibitem{bhat2000finite}
S.~P. Bhat, D.~S. Bernstein, Finite-time stability of continuous autonomous
  systems, SIAM Journal on Control and optimization 38~(3) (2000) 751--766.

\bibitem{polyakov2012nonlinear}
A.~Polyakov, Nonlinear feedback design for fixed-time stabilization of linear
  control systems, IEEE Transactions on Automatic Control 57~(8) (2012)
  2106--2110.

\bibitem{dashkovskiy2011input}
S.~Dashkovskiy, D.~V. Efimov, E.~D. Sontag, Input to state stability and allied
  system properties, Automation and Remote Control 72 (2011) 1579--1614.

\bibitem{sontag2007input}
E.~Sontag, Input to state stability: Basic concepts and results, in: P.~P.
  Nistri, G.~Stefani (Eds.), Nonlinear and Optimal Control Theory,
  Springer-Verlag, Berlin, 2007, pp. 163--220.

\bibitem{efimov2017robust}
D.~Efimov, A.~Polyakov, A.~Levant, W.~Perruquetti, Realization and
  discretization of asymptotically stable homogeneous systems, IEEE
  Transactions on Automatic Control 62~(11) (2017) 5962--5969.

\bibitem{polyakov2019global}
A.~Polyakov, D.~Efimov, B.~Brogliato, Consistent discretization of finite-time
  and fixed-time stable systems, SIAM Journal of Control and Optimization
  57~(1) (2019) 78--103.

\bibitem{shtessel2014sliding}
Y.~Shtessel, C.~Edwards, L.~Fridman, A.~Levant, Sliding Mode Control and
  Observation, Birkhauser, 2014.

\bibitem{efimov2021phase}
D.~Efimov, A.~Polyakov, Finite-time stability tools for control and estimation,
  Foundations and Trends in Systems and Control 9~(2-3) (2021) 171--364.
\newblock \href {http://dx.doi.org/10.1561/2600000026}
  {\path{doi:10.1561/2600000026}}.

\bibitem{kayacan2019feedback}
E.~Kayacan, T.~I. Fossen, Feedback linearization control for systems with
  mismatched uncertainties via disturbance observers, Asian Journal of Control
  21~(3) (2019) 1064--1076.

\bibitem{li2014continuous}
S.~Li, H.~Sun, J.~Yang, X.~Yu, Continuous finite-time output regulation for
  disturbed systems under mismatching condition, IEEE Transactions on Automatic
  Control 60~(1) (2014) 277--282.

\bibitem{moreno2021arbitrary}
J.~A. Moreno, Arbitrary-order fixed-time differentiators, IEEE Transactions on
  Automatic Control 67~(3) (2021) 1543--1549.

\bibitem{sun2014non}
H.~Sun, S.~Li, J.~Yang, L.~Guo, Non-linear disturbance observer-based
  back-stepping control for airbreathing hypersonic vehicles with mismatched
  disturbances, IET Control Theory \& Applications 8~(17) (2014) 1852--1865.

\bibitem{yang2013continuous}
J.~Yang, S.~Li, J.~Su, X.~Yu, Continuous nonsingular terminal sliding mode
  control for systems with mismatched disturbances, Automatica 49~(7) (2013)
  2287--2291.

\bibitem{zhang2018nonsmooth}
C.~Zhang, Y.~Yan, C.~Wen, J.~Yang, H.~Yu, A nonsmooth composite control design
  framework for nonlinear systems with mismatched disturbances: Algorithms and
  experimental tests, IEEE Transactions on Industrial Electronics 65~(11)
  (2018) 8828--8839.

\bibitem{krstic1995nonlinear}
M.~Krstic, I.~Kanellakopoulos, P.~Kokotovic, Nonlinear and Adaptive Control
  Design, Vol.~2, Wiley, New York, 1995.

\bibitem{jiang1994smallgain}
Z.-P. Jiang, A.~R. Teel, L.~Praly, Small-gain theorem for iss systems and
  applications, Mathematics of Control, Signals, and Systems 7 (1994) 95--120.

\bibitem{praly1997generalized}
L.~Praly, Generalized weighted homogeneity and state dependent time scale for
  linear controllable systems, in: 36th IEEE Conference on Decision and
  Control, Vol.~5, IEEE, 1997, pp. 4342--4347.

\bibitem{ye2022robust}
H.~Ye, Y.~Song, Robust adaptive prescribed-time control for parameter-varying
  nonlinear systems, arXiv preprint arXiv:2210.12706.

\bibitem{chitour2020stabilization}
Y.~Chitour, R.~Ushirobira, H.~Bouhemou, Stabilization for a perturbed chain of
  integrators in prescribed time, SIAM Journal on Control and Optimization
  58~(2) (2020) 1022--1048.

\bibitem{holloway2019time}
J.~Holloway, M.~Krstic, Prescribed-time observers for linear systems in
  observer canonical form, IEEE Transactions on Automatic Control 64~(9) (2019)
  3905--3912.

\bibitem{song2017time}
Y.~Song, Y.~Wang, J.~Holloway, M.~Krstic, Time-varying feedback for regulation
  of normal-form nonlinear systems in prescribed finite time, Automatica 83
  (2017) 243--251.

\bibitem{song2018time}
Y.~Song, Y.~Wang, M.~Krstic, Time-varying feedback for stabilization in
  prescribed finite time, International Journal of Robust and Nonlinear Control
  29~(3) (2018) 618--633.

\bibitem{nekhoroshikh2022hyperexponential}
A.~N. Nekhoroshikh, D.~Efimov, A.~Polyakov, W.~Perruquetti, I.~B. Furtat,
  Hyperexponential and fixed-time stability of time-delay systems:
  {L}yapunov--{R}azumikhin method, IEEE Transactions on Automatic Control
  68~(3) (2022) 1862--1869.

\bibitem{wang2024exact}
J.~Wang, K.~Zimenko, A.~Polyakov, D.~Efimov, An exact robust high-order
  differentiator with hyperexponential convergence, IEEE Transactions on
  Automatic Control.

\bibitem{chu2022hyper}
X.~Chu, B.~Zhou, H.~Li, On hyper-exponential stability and stabilization of
  linear systems by bounded linear time-varying controllers, Journal of the
  Franklin Institute 359~(2) (2022) 1194--1214.

\bibitem{efimov2022exact}
D.~Efimov, A.~Polyakov, K.~Zimenko, J.~Wang, An exact robust hyperexponential
  differentiator, in: 2022 IEEE 61st Conference on Decision and Control (CDC),
  IEEE, 2022, pp. 1894--1899.

\bibitem{wang2024hyperexponential}
J.~Wang, K.~Zimenko, D.~Efimov, A.~Polyakov, On hyperexponential stabilization
  of a chain of integrators in continuous and discrete time, Automatica 166
  (2024) 111723.

\bibitem{labbadi2024hyperexponential}
M.~Labbadi, D.~Efimov, Hyperexponential stabilization of double integrator with
  unmatched perturbations, IEEE Control Systems Letters 8 (2024) 2685--2690.
\newblock \href {http://dx.doi.org/10.1109/LCSYS.2024.3510223}
  {\path{doi:10.1109/LCSYS.2024.3510223}}.

\bibitem{alessandri2013time}
A.~Alessandri, A.~Rossi, Time-varying increasing-gain observers for nonlinear
  systems, Automatica 49~(9) (2013) 2845--2852.

\bibitem{efimov2024discretization}
D.~Efimov, Y.~Orlov, Discretization of prescribed-time observers in the
  presence of noises and perturbations, Systems \& Control Letters 188 (2024)
  105820.

\bibitem{aldana2023inherent}
R.~Aldana-L{\'o}pez, R.~Seeber, H.~Haimovich, D.~G{\'o}mez-Guti{\'e}rrez, On
  inherent limitations in robustness and performance for a class of
  prescribed-time algorithms, Automatica 158 (2023) 111284.

\bibitem{deng2024robust}
Y.~Deng, E.~Moulay, V.~L{\'e}chapp{\'e}, Z.~Chen, B.~Liang, F.~Plestan, Robust
  nonsingular predefined-time terminal sliding mode control for perturbed
  chains of integrators, IEEE Transactions on Automatic Control.

\bibitem{alessandri2015increasing}
A.~Alessandri, A.~Rossi, Increasing-gain observers for nonlinear systems:
  Stability and design, Automatica 57 (2015) 180--188.

\bibitem{Brogliato2021}
B.~Brogliato, A.~Polyakov, \href{http://dx.doi.org/10.1002/rnc.5121}{Digital
  implementation of sliding-mode control via the implicit method: A tutorial},
  International Journal of Robust and Nonlinear Control 31~(9) (2021)
  3528--3586.
\newblock \href {http://dx.doi.org/10.1002/rnc.5121}
  {\path{doi:10.1002/rnc.5121}}.
\newline\urlprefix\url{http://dx.doi.org/10.1002/rnc.5121}

\bibitem{Butcher2008}
J.~C. Butcher, Numerical methods for ordinary differential equations, 2nd
  Edition, John Wiley \& Sons, New York, 2008.

\end{thebibliography}

\end{document}